\documentclass[10pt,twocolumn,twoside]{IEEEtran}

\usepackage{amsfonts}
\usepackage{amssymb}
\usepackage{cite}
\usepackage[cmex10]{amsmath}
\usepackage[]{algorithm}
\usepackage{algorithmic}
\usepackage{subeqnarray}
\usepackage{cases}
\usepackage{graphicx}
\usepackage{subcaption}
\usepackage{booktabs}
\usepackage{xcolor}
\usepackage{soul}
\usepackage{caption}
\usepackage{svg}
\usepackage{multirow}
\usepackage{array}
\usepackage{makecell}

\usepackage{xcolor}
\usepackage{xpatch}

\ifCLASSINFOpdf
\else
\fi

\begin{document}
\title{Signal Processing and Learning for\\ Next Generation Multiple Access in 6G}
%
%
%

\author{Wei Chen, \IEEEmembership{Senior Member, IEEE}, Yuanwei Liu, \IEEEmembership{Fellow, IEEE}, Hamid Jafarkhani, \IEEEmembership{Fellow,~IEEE},\\ Yonina C. Eldar \IEEEmembership{Fellow,~IEEE}, Peiying Zhu, \IEEEmembership{Fellow,~IEEE}, Khaled B Letaief,~\IEEEmembership{Fellow,~IEEE}
\thanks{Wei Chen is with the State Key Laboratory of Advanced Rail Autonomous Operation, the School of Electronic and Information Engineering, Beijing Jiaotong University, Beijing, China (email:weich@bjtu.edu.cn).} 
\thanks{Yuanwei Liu is with Queen Mary University of London, U.K (Email: yuanwei.liu@qmul.ac.uk).}
\thanks{Hamid Jafarkhani is with University of California, Irvine, US (Email: hamidj@uci.edu). His work was
supported in part by the NSF Award CCF-2328075.}
\thanks{Yonina C. Eldar is with Weizmann Institute of Science, Israel (Email: yonina.eldar@weizmann.ac.il).}
\thanks{Peiying Zhu is with Huawei Technologies Co., Ltd, Canada (Email: peiying.zhu@huawei.com).}
\thanks{Khaled B Letaief is with The Hong Kong University of Science and Technology, China (Email: eekhaled@ust.hk).}
}

\maketitle
\begin{abstract}
Wireless communication systems to date primarily rely on the orthogonality of resources to facilitate the design and implementation, from user access to data transmission. Emerging applications and scenarios in the sixth generation (6G) wireless systems will require massive connectivity and transmission of a deluge of data, which calls for more flexibility in the design concept that goes beyond orthogonality. Furthermore, recent advances in signal processing and learning, e.g., deep learning, provide promising approaches to deal with complex and previously intractable problems. This article provides an overview of research efforts to date in the field of signal processing and learning for next-generation multiple access, with an emphasis on massive random access and non-orthogonal multiple access. The promising interplay with new technologies and the challenges in learning-based NGMA are discussed.
\end{abstract}

\IEEEpeerreviewmaketitle

\section{Introduction}

\IEEEPARstart{T}{he} fifth generation (5G) wireless systems are currently reaching commercial maturity, and next-generation wireless networks are expected to support extremely high data rates and radically new applications, which require massive connectivity, such as Internet of Everything (IoE), metaverse, augmented reality and industry 4.0. The International Data Corporation (IDC) forecasts that by 2025 there will be 55.7 billion connected devices, most of which will be Internet of Things (IoT) devices, generating 73.1 zettabytes (ZB) of data \cite{IDCforecasts}.


\par
Motivated by these demands, substantial breakthroughs must be achieved by the sixth generation (6G) communication systems \cite{8808168,itu6g}, including i) one terabit per second (Tbps) peak data rate to satisfy the data rate requirements of emerging applications and Augmented Reality (AR) / Virtual Reality (VR) communication, ii) higher spectral efficiency (SE) for massive connectivity, e.g., Internet of Things (IoT) devices that are ten times greater than that of 5G, with low latency and low cost, iii) hyper reliable and low-latency communication, e.g., in an industrial environment for full automation, control, and operation, iv) the support of distributed compute and AI-powered applications, which requires integrated AI and communication, v) the demand for multiple functions, e.g., sensing and navigation \cite{9540344}. To date, wireless communication systems primarily rely on the orthogonality of resources to facilitate design and implementation, from user access to data transmission. The challenging new requirements in 6G call for more flexible design principles that go beyond orthogonality. Especially, breakthroughs in physical-layer random access and multiple-access technologies are critical, in order to provide effective support to all upper-layer services. It is desired to design new techniques beyond the frequency division multiple access (FDMA) in the first generation (1G), time division multiple access (TDMA) in the second generation (2G), and orthogonal frequency division multiple access (OFDMA) in the fourth generation (4G) and 5G. To achieve high SE and high energy efficiency (EE) in the next-generation wireless networks, next generation multiple access (NGMA) technologies will likely deviate from orthogonality as the design principle.


\par
In orthogonal multiple access (OMA) techniques, such as TDMA, FDMA, and OFDMA implemented in 1G through 5G systems, each resource block is allocated to a single user. In contrast to conventional multiple-access techniques, the key concept of NGMA is to intelligently accommodate multiple users in the allotted resource blocks, e.g., time slots, frequency bands, spreading codes and beams, in the most effective manner for different applications. By integrating with other advanced transmission concepts, the performance of multiple access can be further enhanced. Such advanced transmission concepts include RISs \cite{hou2021joint,khaleel2021novel}, cell-free massive MIMO \cite{6798744,8845768,7827017,han2019massive,aref2020deep,li2021scalable}, full duplex relaying, heterogeneous networks, mmWave communications \cite{heath2016overview,xiao2017millimeter,maraqa2020survey}, and THz communications \cite{xu2021graph}. More specifically, by properly integrating NGMA with these advanced concepts, synergy effects can be realized, providing extra benefits for network performance. For example, RISs can provide additional channel paths to build stronger combined channels with apparent strength differences and also artificially re-align users' combined channels to obtain more gain \cite{9316920}. RIS-assisted medium access control protocol is designed, analyzed and optimized in \cite{9693982,9475159,9385372} to improve the system throughput for NGMA. In \cite{10266982}, a joint active and passive beamforming scheme is designed in distributed RIS-aided massive access multiple-input single-output systems with supporting non-orthogonal multiple access (NOMA) and OMA transmissions simultaneously. In cell-free massive MIMO, since access points (APs) serve all users on the network, colliding users located far away can also be detected by some APs based on the distinction of the pilot powers received \cite{10019327}. 

\par
The goals of what people envision being possible within the communication bands are expanding. In particular, there is a goal to use communication signals to perform sensing (or at least assist in sensing). In this case, signal waveform and signal processing are normally jointly designed, e.g., to enable integrated sensing and communication (ISAC) and integrated navigation and communication (INAC). The superposition of the communication and sensing signals in ISAC is a kind of non-orthogonal resource sharing, which shares a similar idea with NOMA. The prominent features of NOMA in efficient interference management and flexible resource allocation inspire sensing interference cancellation for ISAC \cite{9668964,10036107,10024901}. Moreover, our goal in communicating might not be to communicate raw data, but might be to communicate meaning. Recently, with the rapid development of machine learning and artificial intelligence, semantic communications have emerged as a hot research topic, which addresses the semantic- and effectiveness-level communication problems~\cite{Shannon2,Gunduz,10328187,9438648,10225310,qin2021semantic}. Efficient multiple access schemes designed for semantic communication are required to accommodate the new semantic users given the limited radio resources.

\renewcommand\arraystretch{1.2}
\begin{table*}[htp]
\begin{center}
\caption{LIST OF ACRONYMS}
\label{ACRONYMS}
\begin{tabular}{|m{1.5cm}||m{6.7cm}|m{1.5cm}||m{6.7cm}|}
\hline
AA-MF-SIC & Activity-Aware Low-Complexity Multiple Feedback Successive Interference Cancellation & MAP      & Maximum A Posteriori                                     \\
ADMM      & Alternating Direction Multiplier Method                                              & MC-NGMA  & Multi-Concept-Oriented Next Generation Multiple Access    \\
AI        & Artificial Intelligence                                                              & MF-NGMA  & Multi-Functional-Oriented Next Generation Multiple Access \\
AMP       & Approximate Message Passing                                                          & MIMO     & Multiple-Input Multiple-Output                            \\
ANOMA     & Asynchronous Non-Orthogonal Multiple Access                                          & ML       & Machine Learning                                          \\
AP        & access point                                                                         & MLE      & Maximum Likelihood Estimation                             \\
AR        & Augmented Reality                                                                    & MMSE     & Minimum Mean Squared Error                                \\
AUD       & Active User Detection                                                                & MMTC     & Massive Machine-Type Communication                        \\
AWGN      & Additive White Gaussian Noise                                                        & MMV      & Multiple Measurement Vector                               \\
BC        & Broadcast Channel                                                                    & mmWave   & Millimeter-wave                                           \\
BCS       & Bayesian Compressive Sensing                                                         & MnAC     & Many-Access Channel                                       \\
BER       & Bit Error Rate                                                                       & MPA      & Message Passing Algorithm                                 \\
BS        & Base Station                                                                         & MRA      & Massive Random Access                                     \\
BOMP      & Block Orthogonal Matching Pursuit                                                    & MRC      & Maximum Ratio Combining                                   \\
BSAMP-CP  & Backward Sparsity Adaptive Matching Pursuit with Checking and Projecting             & MS       & Mode Switching                                            \\
BSBL      & block sparse Bayesian learning                                                       & M-SP     & Modified Subspace Pursuit                                 \\
CCS       & Code Compressed sensing                                                              & MT-NGMA  & Multi-Tool-Oriented Next Generation Multiple Access       \\
CDMA      & Code Division Multiple Access                                                        & NFC      & Near-Field Communications                                 \\
CD-NOMA   & Code-Domain Non-Orthogonal Multiple Access                                           & NGMA     & Next Generation Multiple Access                           \\
CE        & Channel Estimation                                                                   & NN       & Neural Network                                            \\
CF mMIMO  & Cell-Free massive Multiple Input Multiple Output                                     & NOMA     & Non-Orthogonal Multiple Access                            \\
C-NOMA    & Cooperative Non-Orthogonal Multiple Access                                           & OAMP     & Orthogonal Approximate Message Passing                    \\
CRAN      & Cloud Radio Access Network                                                           & OFDMA    & Orthogonal Frequency Division Multiple Access             \\
CS        & Compressed Sensing                                                                   & OLS      & Orthogonal Least Square                                   \\
CSA       & Coded Slotted Aloha                                                                  & OMA      & Orthogonal Multiple Access                                \\
CSI       & Channel State Information                                                            & OMP      & Orthogonal Matching Pursuit                               \\
CS-MUD    & Compressive Sensing based Multi-User Detection                                       & PDMA     & Pattern Division Multiple Access                          \\
CTSMA     & Coded Tandem Spread Spectrum Multiple Access                                         & PD-NOMA  & Power-Domain Non-Orthogonal Multiple Access               \\
DDPG      & Deep Deterministic Policy Gradient                                                   & PtrNet   & Pointer Network                                           \\
DL        & Deep Learning                                                                        & PUPE     & Per-User Probability of Error                             \\
DL        & Downlink                                                                             & RAN      & Radio Access Networks                                     \\
DNN       & Deep Neural Network                                                                  & RAR      & Random Access Response                                    \\
DoFs      & Degrees-of-Freedom                                                                   & RIS      & Reconfigurable Intelligent Surface                        \\
DQN       & Deep Q-Network                                                                       & RL       & Reinforcement Learning                                    \\
DRL       & Deep Reinforcement Learning                                                          & SBL      & Sparse Bayesian Learning                                  \\
EE        & Energy Efficiency                                                                    & SC       & Spatial Coupling                                          \\
ELAA      & Extremely Large-scale Antenna Array                                                  & SCA      & Successive Convex Approximation                           \\
ES        & Energy Splitting                                                                     & SCMA     & Sparse Code Multiple Access                               \\
FDMA      & Frequency Division Multiple Access                                                   & SDMA     & Space Division Multiple Access                            \\
FISTA     & Fast Iterative Shrinkage-Thresholding Algorithm                                      & SE       & Spectral Efficiency                                       \\
FMCW      & Frequency Modulated Continuous Wave                                                  & SI       & Side Information                                          \\
GGAMP     & Gaussian Generalized Approximate Message Passing                                     & SIC      & Successive Interference Cancellation                      \\
GGSO      & Group Gram-Schmidt Orthogonalization                                                 & SISD     & Structured Iterative Support Detection                    \\
GOMP      & Group Orthogonal Matching Pursuit                                                    & SISO     & Single-Input Single-Output                                \\
HTC       & Human-Type Communication                                                             & SMV      & Single Measurement Vector                                 \\
ICT       & Information and Communication Technology                                             & SNR      & Signal-to-Noise Ratio                                     \\
IDC       & International Data Corporation                                                       & SSM      & Structured Sparse Model                                   \\
INAC      & Integrated Navigation And Communication                                              & STARS    & Simultaneously Transmitting And Reflecting Surfaces       \\
IoE       & Internet of Everything                                                               & T\&R     & Transmission and Reflection                               \\
IoT       & Internet of Things                                                                   & TA-BSASP & Threshold Aided Block Sparsity Adaptive Subspace Pursuit  \\
ISAC      & Integrated Sensing And Communication                                                 & TBM      & Tensor-Based Modulation                                   \\
ISI       & Inter-Symbol Interference                                                            & TDMA     & Time Division Multiple Access                             \\
ISTA      & Iterative Shrinkage-Thresholding Algorithm                                           & THz      & Terahertz                                                 \\
IUI       & Inter-User Interference                                                              & TS       & Time Switching                                            \\
LDMA      & Location-Division Multiple Access                                                    & UAV      & Unmanned Aerial Vehicle                                   \\
LDPC      & Low-Density Parity-Check                                                             & UL       & Uplink                                                    \\
LDS-CDMA  & Low-Density Signature Code Division Multiple Access                                  & USL      & Unsupervised Learning                                     \\
LIP       & Linear Inverse Problem                                                               & VAE      & Variational Autoencoder                                   \\
LoS       & Line of Sight                                                                        & VAMP     & Vector Approximate Message Passing                        \\
LSTM      & Long Short-Term Memory                                                               & VR       & Virtual Reality                                           \\
MAC       & Multiple-Access Channel                                                              & ZB       & Zettabytes                                                \\ \hline
\end{tabular}
\end{center}
\end{table*}
\renewcommand\arraystretch{1}

\par
Massive random access (MRA) and NOMA\footnote{In this paper we will use the acronym ``NOMA" to denote the class of non-orthogonal communication techniques that employ a single transmission codebook per user and SIC at the receiver. Some other classes of non-orthogonal techniques use multiple codebooks for each user, including time-sharing schemes \cite{cover2006elements} and rate-splitting schemes \cite{9831440,10038476,9451194}.} are key technologies in NGMA. Random access is a necessary process for establishing wireless links between terminals and networks, which is extremely challenging in the scenario of a large number of devices, i.e., MRA. On the other hand, by employing superposition coding \cite{cover2006elements} at the transmitter and successive interference cancellation (SIC) \cite{4276942} at the receiver, NOMA allows more users to be served than conventional multiple access techniques. Advances in computing and digital signal processing technologies now enable us to deal with some complex and previously intractable problems arising in NGMA \cite{9903376,9605579,9252937,8827912,9875174}. Recent research has shown improvements in connectivity \cite{9903376,9605579,9252937}, system capacity (and SE) \cite{gui2018deep,8827912,saetan2019application} and reduction in overall latency \cite{ye2019deep,fu2019dynamic}. In particular, artificial intelligence and machine learning (AI / ML) have become one of the main flagship activities in the 3rd generation partnership project (3GPP), which includes a wider use of AI / ML for optimizing radio access networks (RAN) and air interfaces \cite{9606720,10367817,10445236}. AI-empowered multiple access for spectrum sensing, protocol designs, and optimizations is reviewed in \cite{2406.13335}. The mining, classification, recognition, prediction, and learning capabilities of advanced signal processing technology make wireless networks more intelligent, and it is also essential to study the implementative and interpretative nature of technology to improve the performance of the NGMA. 

\par
The organization of this article is as follows. Section II reviews the background of MRA and NOMA for NGMA, including both fundamental limits and practical schemes. We provide an overview of the current research contributions using advanced signal processing and machine learning for MRA and NOMA in Section III and Section IV, respectively. The interplay between NGMA and other new next-generation technologies and scenarios, e.g., age of information (AoI), near-field communications (NFC), ISAC, simultaneously transmitting and reflecting surfaces (STARS) and semantic communications, is discussed in Section V. In Section VI, we highlight several implementation challenges in using learning-based methods for NGMA. Finally, Section VII concludes the paper. Table \ref{ACRONYMS} provides a list of acronyms used in this paper.

\begin{figure*}[!t]%
    \centering
    \includegraphics[width=0.9\textwidth]{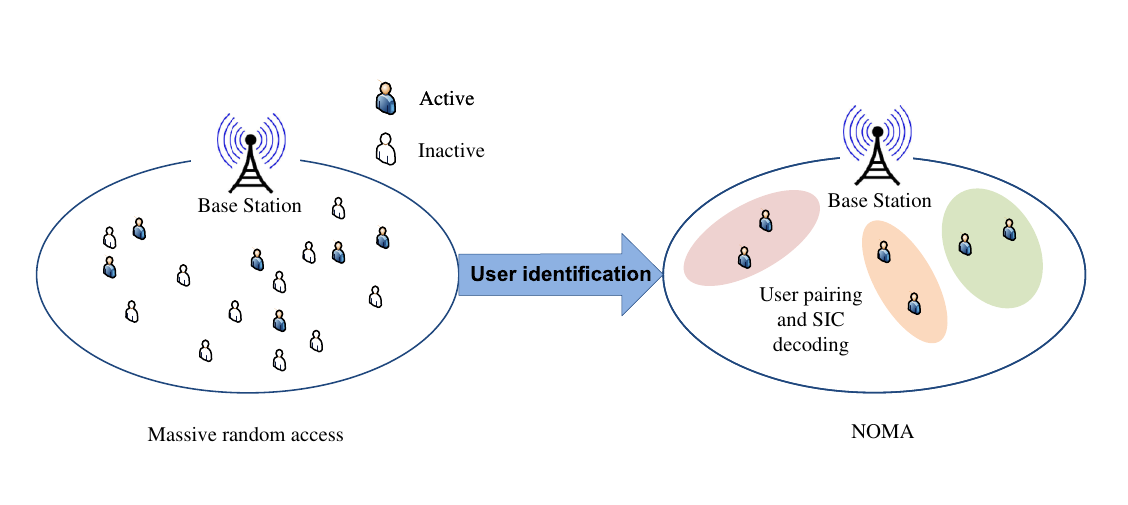}
    \caption{Illustration of the MRA and NOMA scenarios.}
    \label{MA-NOMA}
\end{figure*}

\section{Background of Massive Random Access and Non-Orthogonal Multiple Access}
In this section, we introduce the background of MRA and NOMA, with the focus on the non-orthogonal use of wireless resources to enhance massive connectivity and data transmission. 

\subsection{Massive Random Access}
Massive machine-type communication (MMTC) is one of the three major application scenarios in 5G. It is still of significant importance in Beyond 5G (B5G) and 6G, making it one of the most popular research directions in recent years. Conventional communication systems are mainly designed for human-type communication (HTC), with the goal of supporting the transmission of a large amount of data for a small number of users at high data rates. MMTC has distinct characteristics from existing HTCs, such as uplink dominant, massive user access, sporadic communication, small packet transmission, low communication rate and low cost devices. The indicator for access devices in 5G is $10^6$ users per square kilometer, while the number of devices that need to be supported in 6G has been increased to $10^7$ users per square kilometer \cite{8766143}. If the number of users is large, there are issues related to pilot collision for reliable channel estimation. Some NOMA schemes have been extensively studied in 5G standards \cite{chen2018toward}. This is still a huge challenge for the design of next-generation communication systems, as the existing access mechanism for 5G MMTC is not sufficient to efficiently and reliably support the transmission of small data packets for massive devices. As shown in Fig. \ref{MA-NOMA}, the goal of MRA is to detect a few active devices from a large number of devices, and then NOMA can be used to support those multiple active devices. Due to the complexity of SIC in NOMA, it is advisable to divide users into groups with orthogonal resources, and apply NOMA for each group independently.

\par
Various coordinated multiple access technologies, e.g., FDMA, TDMA, CDMA\footnote{Here we mean the synchronous CDMA case.}, OFDMA, space division multiple access (SDMA), and NOMA, require specific protocols to efficiently manage communication among users with access to the same systems, whose fundamental limits are usually characterized as the classical multiple-access channel (MAC) capacity. With an increasing number of randomly activated users, the overhead of coordinating active users would overwhelm the system. Uncoordinated multiple access allows a set of users to transmit over a common wireless medium opportunistically and independently, resulting in different theoretical analyses. 

\subsubsection{Basic Principles}
From the perspective of information theory, the fundamental limits of MRA have been characterized in various settings. The many-access channel (MnAC) is studied in \cite{7852531}, which provides the asymptotic regime where the total number of users $n$ increases as the blocklength $m$ tends to infinity. Given the stability requirement, the total number of users $n$ increases with the frame length \cite{7513390,9174915}. In \cite{10042422}, the age of information for random access is derived, revealing its relation with the number of total nodes and their activation probability. Finite-blocklength achievability bounds for the random access channel under average-error and maximal-power constraints are presented in \cite{9535162}. Non-asymptotic random coding achievability bounds with the criterion of per-user probability of error (PUPE) are derived \cite{8006984,8849764}. Under the PUPE
criterion, non-asymptotic achievability and converse bounds on the minimum required energy per bit for MRA in MIMO quasi-static Rayleigh fading channels are established. 

\par
A fundamental problem in MRA is to detect the activity of users. Assume that there are in total $n$ users sharing a wireless channel, and both the transmitter and the receiver have a single antenna. Each user can be assigned a unique dedicated codeword represented by an $m$-dimensional vector in $\mathbf{A}$, leading to a way of coordinated multiple access. Alternatively, user could randomly select codewords from the codebook $\mathbf{A}$ to transmit, leading to uncoordinated multiple access. The signal at the BS contributed by codeword $\mathbf{a}_j$ can be modeled as $x_j\mathbf{a}_j$ where $x_j$ is a complex scalar that represents the product of the transmitted symbol and channel gain. If the codeword is not transmitted, $x_j=0$. The total received signal at the BS can be expressed as
\begin{equation}\label{eq:CS}
\mathbf{y} =\sum_{j=1}^n x_j\mathbf{a}_j + \mathbf{z} 
=\mathbf{Ax}+\mathbf{z}  ,
\end{equation}
where $\mathbf{z}$ represents noise and $\mathbf{A}=[\mathbf{a}_1,\ldots,\mathbf{a}_n]$ denotes the codebook. Note that this is a general mathematical model, and can be easily solved if $m\geq n$. In the case of users with unique codewords, active users and their corresponding channel are obtained according to the estimated $\mathbf{x}$. For users randomly selecting codewords from the common codebook, some collisions may occur if multiple users select the same codeword, and further collision resolution techniques are required to correctly identify all access users. When $m\ll n$, i.e., the total number of observations is much smaller than the total number of codewords, it is not possible to solve for a unique $\mathbf{x}$ even if the noise term is equal to zero, as there is a large number of solutions satisfying equation (\ref{eq:CS}). We are interested in this case, as it is desired to support a large number of potential users with limited resources. Since most codewords are inactive, a suitable sparsity constraint on $\mathbf{x}$ may rule out all solutions except the one that is expected. Therefore, the activity detection problem in MRA is equivalent to the sparse recovery problem in compressed sensing (CS), and the identities of active users are obtained according to the activated codewords. Furthermore, with some variation of assumptions, different problems in MRA boil down to the same sparse recovery problem in equation (\ref{eq:CS}). For example,
\begin{itemize}
  \item if all users transmit with the unit power, recovering $\mathbf{x}$ in equation (\ref{eq:CS}) leads to joint activity detection and channel estimation; 
  \item for the Gaussian MAC channel, the message of user $j$ can be carried by $x_j$ and recovering $\mathbf{x}$ decodes the messages of active users. 
\end{itemize}
Searching for an $\mathbf{x}$ in equation (\ref{eq:CS}), that has these sparsity properties, leads to the problem of minimizing the number of selected codewords that could have generated the received signal $\mathbf{y}$, which involves minimizing the ``0-semi-norm" of $\mathbf{x}$. It is an ill-posed sparse linear inverse problem, which has a non-continuous objective function and is NP-hard \cite{BAI2020107729}. An alternative approach is to employ a smoothed objective function, e.g., the $\ell_1$ norm of $\mathbf{x}$, and the CS theory claims that $\mathbf{x}$ can be recovered with a very high possibility when the length of the codeword satisfies 
\begin{equation}\label{eq:CS-limit}
m=\mathcal{O}\left(s\log \frac{n}{s}\right) ,
\end{equation}
where $s$ denotes the number of active users. The above results assume a single BS antenna. Multiple antennas at the BS can bring additional improvements in MRA \cite{8323218,9374476}. In addition to these CS-based methods, structured access protocols, e.g., coded tandem spread spectrum multiple access (CTSMA) \cite{ma2018coded}, which solve the access problem from the coding perspective, have also attracted considerable attention.

\begin{figure*}[!t]%
    \centering
    \includegraphics[width=0.9\textwidth]{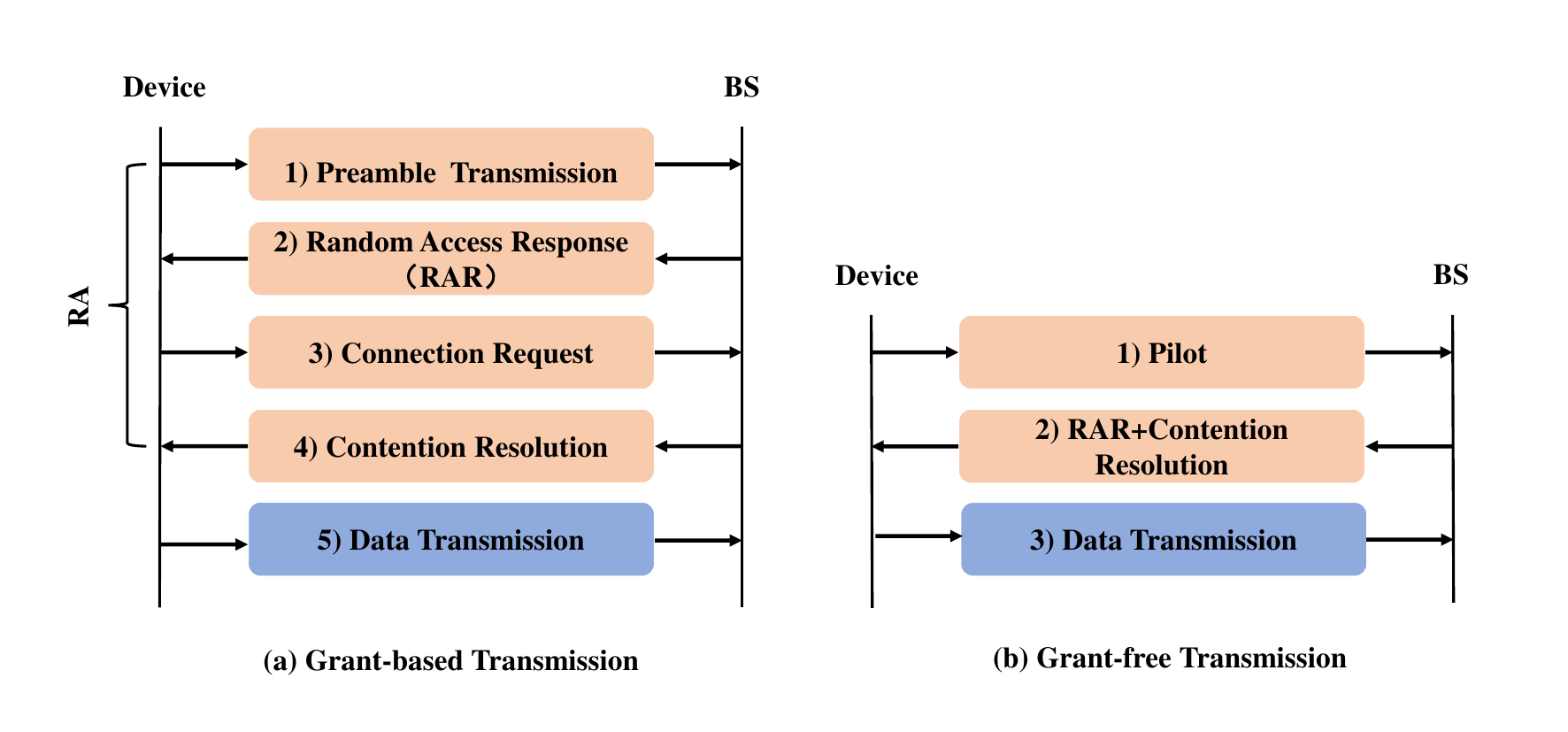}
    \caption{Grant-base random access and grant-free random access schemes.}
    \label{GB-GBF}
\end{figure*}

\par
From a signal processing standpoint, there are two primary approaches to improving the performance of MRA. One is to employ advanced signal processing techniques that extract and leverage various prior information or structures embedded in the signal. More details will be presented in Section III.A. Another way is to employ the power of data-driven methods, or namely, machine learning methods. Plenty of empirical evidence has shown that deep learning can significantly increase convergence speed and accuracy in solving optimization problems, and even exploit knowledge embedded in complex problems that is usually difficult to model. More research results along this line will be presented in Section III.B.

\subsubsection{Grant-Free Random Access}
Most of the wireless communication systems currently in operation use a grant-based random access mechanism. Each active device utilizes a preamble sequence to notify the BS that it has become active, after which the BS sends a response in the form of a grant for transmitting the device's data. Therefore, active devices must engage in multiple signaling interactions with the BS to establish connections and allocate specific resources for data transmission. However, due to the majority of small data packets transmitted in MMTC, the negligible signaling overhead in large data packet transmission accounts for a relatively large proportion in MMTC, which cannot be ignored and often leads to an excessive proportion of signaling overhead, resulting in low EE and SE for the transmission of small data packets. Furthermore, the cumbersome handshake process in grant-based random access inevitably leads to significant latency, which makes the grant-based mechanism unsuitable for low latency applications.

\par
A promising technology to support MMTC is grant-free random access, where channel resources are accessed by active devices without obtaining permissions from the BS. The grant-base random access and grant-free random access schemes are illustrated in Fig. \ref{GB-GBF}. Grant-free random access brings significant benefits in reducing the signaling overhead and access latency. It has the potential to be applied in many emerging scenarios. A comprehensive review of grant-free random access protocols and techniques in satellite-based IoT networks is showed in \cite{10492466}. A typical grant-free random access scheme is to assign a unique codeword, also called preamble, pilot or signature in the literature, to each device. To support massive devices with limited resources, non-orthogonal codewords need to be employed. Active devices initiate access by sending their codewords, and the BS conducts activity detection via detecting these codewords from the received signal. Another type of grant-free random access is unsourced random access, also called uncoordinated multiple access, in which all devices share the same codebook to transmit messages directly and the BS does not have the identity information of the active device. The goal of unsourced random access is to decode the information transmitted by the devices. If the device needs to be identified, the identity information can be included in the transmitted message. The benefits brought by grant-free random access comes at an increased likelihood of collisions resulting from uncoordinated channel access and inter-user interference due to the non-orthogonal access resources. Therefore, novel reliability enhancement techniques are needed. 

\subsection{Non-Orthogonal Multiple Access}
In OMA, a single wireless resource can be allocated to only one user, such as by frequency division or time division. However, with the development of 5G, B5G, and the arrival of 6G, there will be an explosive increase in data traffic and access demand and traditional OMA cannot meet a series of challenging indicators, such as connectivity density and spectral efficiency. NOMA is proposed to improve spectrum efficiency and access capacity and may encompass a variety of use cases or deployment scenarios \cite{chen2018toward}. Compared to OMA, different NOMA users are multiplexed without the constraint of strict orthogonal radio resource allocation. With the help of optimized resource allocation and advanced receivers, the overall throughput of the system can be greatly improved.
\begin{figure*}[!t]%
    \centering
    \includegraphics[width=0.9\textwidth]{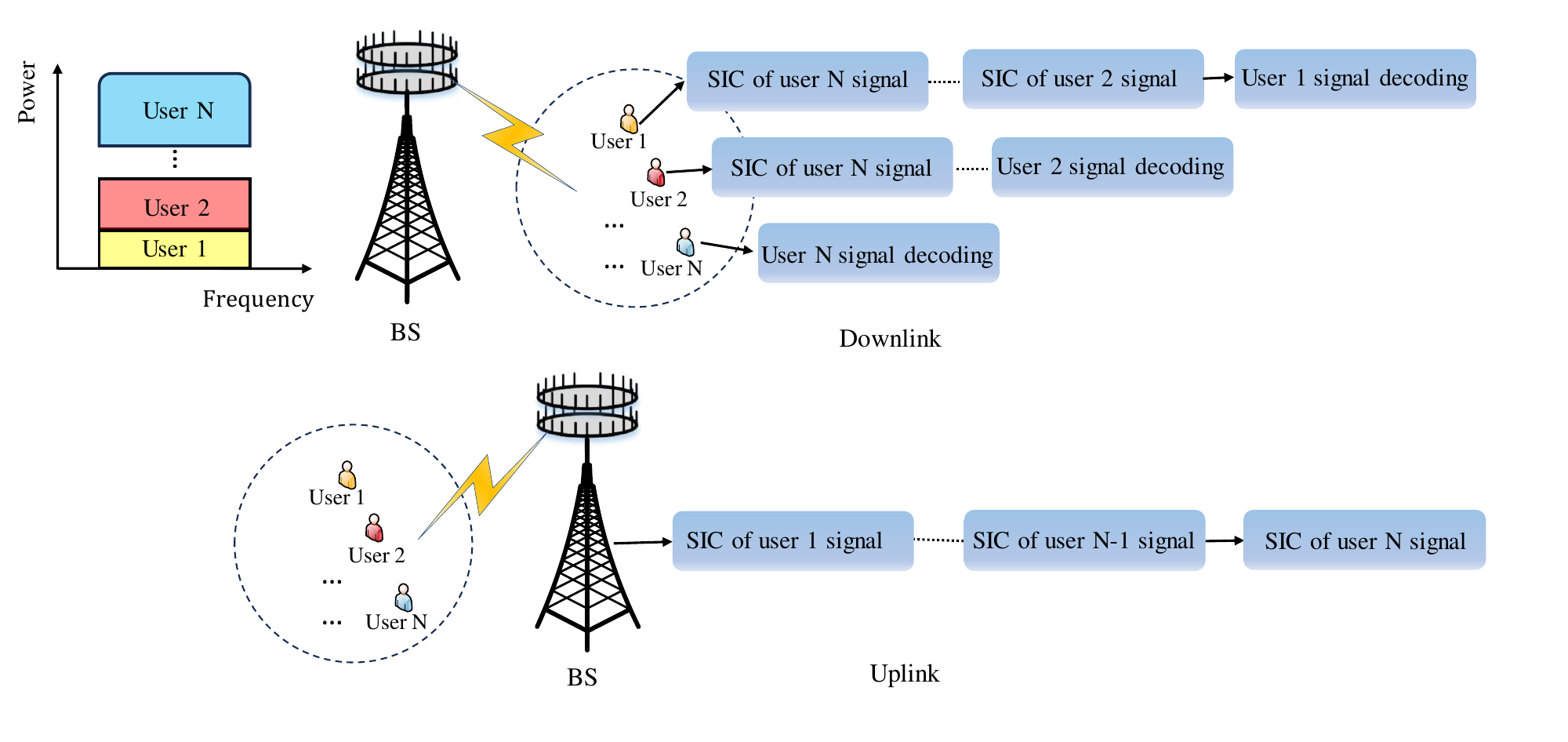}
    \caption{An illustration of downlink and uplink PD-NOMA.}
    \label{PD-NOMA}
\end{figure*}
\subsubsection{Basic Principles}
The capacity region of the multiaccess fading uplink channel is characterized in \cite{737513,737514}. MolavianJazi and Laneman establish the second-order achievable rate regions for the Gaussian MAC in \cite{7300429}, and Yavas et al. give the finite-blocklength achievability bounds for the Gaussian MAC and random access channel (RAC) under average-error and maximal-power constraints in \cite{9535162}. The achievable rate region of OMA is a subset of the capacity region for single-input single-output (SISO) Gaussian MACs and broadcast channels (BCs), since each user can only use a portion of the resource blocks in OMA \cite{cover2006elements}. NOMA can achieve reliable communication at Synchronous a larger set of rates in the SISO cases. Consider synchronous power-domain NOMA (PD-NOMA), where users' signals are superimposed via superposition coding and then receivers adopt SIC techniques for interference cancellation. Fig. \ref{PD-NOMA} illustrates the downlink and uplink signal transmission and processing of PD-NOMA. For downlink transmission with two users, the transmitted signal of the BS using superposition coding can be expressed as
\begin{equation}
x = \sqrt {{p_1}}{s_1} + \sqrt {{p_2}} {s_2},
\label{1}
\end{equation}
where ${s_1}$ and ${s_2}$ are the transmitted signals for the two users, and ${p_1}$ and ${p_2}$ are the corresponding transmit power. Let ${h_1}$ and ${h_2}$ denote the channel coefficients of User 1 and User 2, respectively, and $\sigma^2$ denote the noise power. Under the assumption of ${\left| {{h_1}} \right|^2} > {\left| {{h_2}} \right|^2}$, User 1 first decodes the signal of User 2 and subtracts it from the received signal through SIC and then decodes its intended signal. On the other hand, User 2 directly decodes its intended signal by treating User 1’s signal as noise. Since User 1's channel is stronger, User 1's achievable rate is larger than that of User 2 and User 1 can decode User 2's signal and cancel it. Therefore, the achievable rates of User 1 and User 2 are given by
\begin{equation}
{R_1} = {\log _2}\left(1 + \frac{{{p_1}{{\left| {{h_1}} \right|}^2}}}{{{\sigma ^2}}}\right)
\end{equation}
and
\begin{equation}
{R_2} = {\log _2}\left(1 + \frac{{{p_2}{{\left| {{h_2}} \right|}^2}}}{{{p_1}{{\left| {{h_2}} \right|}^2} + {\sigma ^2}}}\right),
\end{equation}
respectively. For synchronous uplink NOMA transmission, the two users send their signals to the BS. At the BS, users 1' signal which experiences stronger channel is first decoded and then users 2' signal. In both the downlink and the uplink, OMA is only able to achieve some specific points on the unicast capacity region, e.g., the points where only one user is active, while NOMA is more flexible \cite{cover1972broadcast}. For example, NOMA with joint decoding achieves the ``corner points" of the multiple access region in the two-user single-antenna uplink case, and points on the ``dominant face" can be achieved by time sharing between the two ``NOMA with sequential decoding" schemes or rate splitting \cite{485709}. For the Gaussian single antenna downlink channel, the standard superposition coding approach is indeed a NOMA scheme, and can achieve an arbitrary point on the boundary of the unicast capacity region. For the multiple antenna case, the channel is not degraded, and one needs a more sophisticated form of transmission, e.g., dirty-paper coding \cite{1683918}.

\par
Although NOMA has advantages such as high bandwidth efficiency and connectivity, there are still many challenges to be tackled. As mentioned above, SIC is the key technology for demodulating superimposed signals. However, if errors occur during the demodulation of weak user's signals in the first stage, it will subtract the wrong modulated symbol from the received superimposed signal and cause a decrease in the performance of the demodulation in the second stage. Accurate channel state information (CSI) is crucial in NOMA. Error in estimating the CSI may lead to user grouping, ordering, and decoding errors. The increasing number of connected users and larger scale antennas in the next generation wireless networks will make the acquisition of CSI more challenging. Another challenge in NOMA is the security issue. Existing research on secure communication to avoid untrusted users and external eavesdroppers usually assumes the availability of full, partial, or statistical CSI. However, it is difficult to obtain CSI in the case of passive eavesdropping. Some interesting techniques, e.g., multi-dimensional directional modulation, cyclic feature suppression-based techniques, and channel-based inter-leaving may be potential solutions for enhancing security \cite{wu2018survey}.

\subsubsection{Asynchronous NOMA}
Because of the distributed nature of multi-user networks and the effects of multipath and propagation delays, signals from different users experience different time delays to the receiver \cite{MGHJ16}. Perfect symbol synchronization requires feedback and coordination, which complicates the system greatly. Having multiple antennas at the receiver makes it even more complicated. Assuming a multiple antenna receiver in uplink NOMA, one can only synchronize the symbols of the received signals perfectly at one of the receive antennas and the rest of antennas may experience imperfect timing synchronization among received signals. Even if such a complete symbol synchronization is possible, it is not clear if it is desirable.
In fact, the MAC’s capacity-region for a system with time asynchrony and rectangular waveforms in CDMA is larger than that of a perfectly synchronized system \cite{Verdu89}. Also, intentionally adding symbol-level timing offsets to make the signals asynchronous has shown advantages in managing interference in MIMO \cite{MGHJ16,Shao2007,DasRao11} and 
in differential decoding for MACs \cite{SPHJ15,MAHJ15}. The possible advantages along with the difficulties in achieving perfect synchronization motivate a thorough analysis and design of NOMA systems under imperfect timing synchronization.

\par
For uplink asynchronous NOMA (ANOMA), the capacity-region for two users utilizing different pulse shaping methods has been derived in \cite{MGXZHJ21}. The analysis shows that the asynchronous transmission enlarges the capacity-region. In fact, not only does asynchronous transmission,
i.e., timing mismatch, improve the performance of NOMA systems, but also SIC is not optimal for asynchronous NOMA \cite{MGXZHJ21}. Let us assume User $k$ transmits  $\sqrt{P_k}s_k[n]$, where $s_k[n]$ denotes the $n$th normalized
transmitted symbol and $P_k$ denotes the transmit power. Then, under imperfect timing synchronization among different users, the received signal at the BS is given by
\begin{flalign}
y(t)=\sum_{n=1}^N\sum_{k=1}^K{h_k \sqrt{P_k} s_k[n]  p(t-nT-\tau_k T)+\eta(t)},
\end{flalign}
where $N$ represents the frame length, $K$ is the number of users, $T$ is the symbol interval, $\tau_k$ denotes the relative time delay at User $k$, $p(.)$ is the pulse-shaping filter, and $\eta(t)$ denotes
the normalized additive white Gaussian noise (AWGN).
The set of sufficient statistics can be found by proper filtering at the receiver and over-sampling $K$ times, each time synched with one of the users \cite{MGHJ16}. The resulting input-output model is
\begin{flalign}
\label{system_model}
\boldsymbol{y}=\boldsymbol{R}\boldsymbol{H}\boldsymbol{s}+\boldsymbol{\eta},
\end{flalign}
where $\boldsymbol{y}, \boldsymbol{R}, \boldsymbol{H}, \boldsymbol{s}$, and $\boldsymbol{n}$ represent the set of samples, the timing offsets matrix, the effective channel matrix, the transmited symbols, and the noise vector, respectively. Note that the perfect synchronization results in the conventional system model of $y[n]=\sum_{k=1}^K h_k \sqrt{P_k} s_k[n]+n[n]$. 

\par
In a perfectly symbol-synchronized NOMA system, at each time instant, only the inter-user interference (IUI) degrades the performance. 
    In an asynchronous NOMA, not only IUI but also inter-symbol interference (ISI) degrades the performance. Therefore, the conventional 
    SIC is not optimal anymore and efficient joint detection methods may be needed. In addition, because of the timing asynchrony, sufficient statistics results in over-sampling and the corresponding sampling diversity \cite{MGHJ16} can improve the overall performance.
    While the conventional wisdom suggests that 
    the imperfect timing increases the overall interference and the overall performance is degraded, surprisingly, imperfect timing in fact decreases the overall interference, because of the benefits of sampling diversity \cite{MGXZHJ21}, as the reduction in IUI outweighs the addition of ISI \cite{XZBHHJ19}.

\par
For downlink asynchronous NOMA, the superposition is performed at the transmitter and the transmitted signal is received by the intended users \cite{cui2017asynchronous,MGHJwcnc19}. Adding intentional time delays at the transmitter can in fact improve the performance. The transmitted signal, including the added intentional time delays, can be written as $s(t)=\sum_{n=1}^N\sum_{k=1}^K{\sqrt{P_k}s_k[n]p(t-nT-\tau_k T)}$, and the received signal by User $k$ is represented as
\begin{flalign}
y_k(t)=h_ks(t)+\eta_k(t).
\end{flalign}
Similar to the case of uplink ANOMA, the set of sufficient statistics can be found by proper filtering and over-sampling as
\begin{flalign}
\label{system_model2}
\boldsymbol{y_k}=h_k \boldsymbol{R_k}\boldsymbol{s}+\boldsymbol{\eta_k},
\end{flalign}
where $\boldsymbol{y_k}, \boldsymbol{R_k}, h_k$, and $\boldsymbol{\eta_k}$ represent the set of samples, the timing offsets matrix, the effective channel coefficient, and the noise vector at User $k$, respectively. Note that the perfect synchronization results in the conventional system model of $y_k[n]=h_k\sum_{j=1}^K \sqrt{P_j}s_j[n]+n_k[n]$.
Note that any User $j$ with $|h_j|<|h_k|$ will have a smaller capacity compared with that of User $k$. Therefore, its transmission rate will be chosen lower than the capacity of User $k$ and from an information theoretical point of view, User $K$ can decode the signal of a weaker user with no error. Using SIC, each user decodes all signals from weaker users and removes them from the received signal. Then, it considers the remaining interference from the stronger users as noise and decodes its own message. Downlink ANOMA has to deal with similar imperfect timing issues discussed for uplink ANOMA; however, it can perform coordinated resource management. For instance, power allocation can be performed globally with collective CSI from all users. More importantly, because of the virtual-MIMO nature of the system model in  (\ref{system_model2}), enhanced joint processing can be performed at the transmitter to exploit the virtual-MIMO structure. The joint processing at the transmitter can be realized by intelligent correlated precoding and circumvents the need for SIC, as done in referece \cite{MGHJwcnc19}. The idea of adding intentional time delays have been extended to intentionally adding 
frequency offsets in NOMA multi-carrier and OFDM systems \cite{8669872,10056854}.

\section{Massive Random Access}
In this section, we introduce model-based methods, using advanced signal processing, and learning empowered methods for MRA. 

\begin{figure*}
    \centering
    \includegraphics[width=0.9\textwidth]{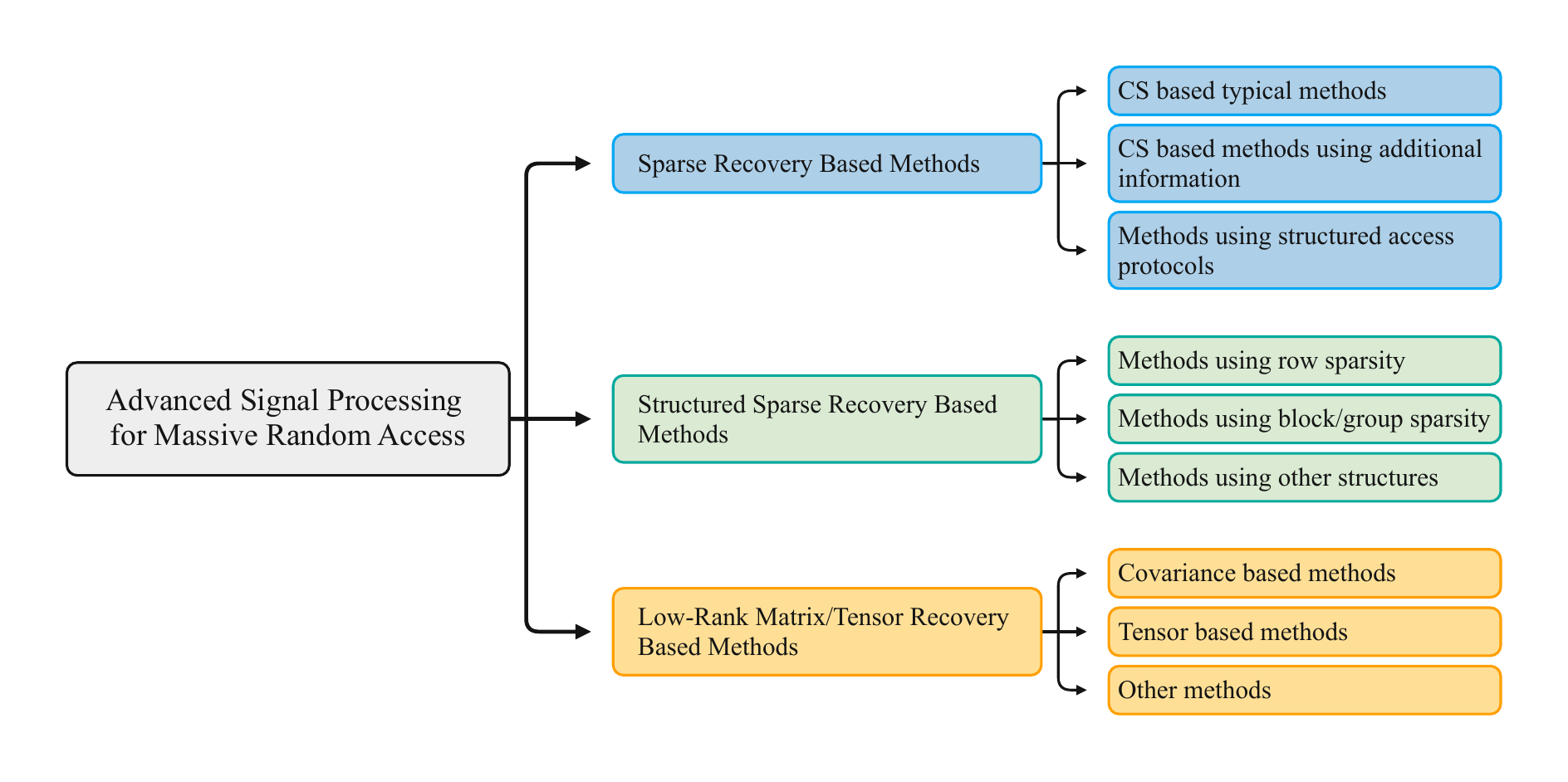}
    \caption{Advanced signal processing for MRA.}
    \label{SP4MRA}
\end{figure*}

\subsection{Model-Based Methods for MRA}
For active user detection in MRA, model- methods are classified into sparse recovery based methods, structured sparse recovery based methods, and low-rank matrix/tensor recovery based methods, as shown in Fig. \ref{SP4MRA}. 

\subsubsection{Sparse Recovery Based Methods for MRA}
\par
As discussed in the previous section, the grant-free access scheme is attractive in MMTC, where active devices directly transmit their pilot and data to the BS without grant. Simultaneously conducting accurate active user detection (AUD) and CE with non-orthogonal pilots is critically important and challenging for device to successfully establish reliable communication with the BS. Fortunately, the sporadic nature of MMTC traffic results in sparse user activity patterns. By using sparsity, the non-orthogonal pilot detection and CE problem can be modeled as a sparse linear inverse problem (LIP), thus reminiscent of the application of CS. By solving the LIP, not only do we realize highly reliable AUD and accurate CE, but also we alleviate the multi-user interference problem. 

\textbf{CS Based Typical Methods:}
Various CS based algorithms could be used to solve the AUD and CE problem, e.g., iterative shrinkage-thresholding algorithm (ISTA) \cite{daubechies2004iterative}, fast ISTA (FISTA) \cite{WOS:000278101000009}, orthogonal matching pursuit (OMP) \cite{4385788}, message passing \cite{WOS:000271637500010}, and sparse Bayesian learning (SBL) \cite{8074806,8039502}. 
\begin{itemize}
\item \textit{OMP}: Schepker and Dekorsy used OMP \cite{tropp2007signal} and orthogonal least squares (OLS) for MUD, which iteratively selected the most probable active users and subsequently estimated their data \cite{schepker2011sparse}. Group orthogonal matching pursuit (GOMP) was further exploited to deal with the asynchronicity case, where chips are not received at the same time \cite{bockelmann2013compressive}. As CS based methods have high computational complexity, most literature considers $\leq 10^4$ devices in their experiments. To achieve good scalability, a contention-based MRA scheme was proposed in \cite{9148341} to support massive access, e.g., $\geq 10^6$, within a certain time-frequency resource, and a one-step access scheme was adopted to save energy and spectrum resources.

\item \textit{AMP}: The approximate message passing algorithm
(AMP) \cite{WOS:000271637500010} is attractive in AUD and CE owing to its high computational efficiency. Reference \cite{8264818} considered two cases depending on whether the large-scale component of channel fading is known in the BS and designed the minimum mean squared error (MMSE) denoiser for AMP according to the channel statistics. It also provided an analytical characterization of the probabilities of false alarm and missed detection through state evolution. Although AMP is an iterative algorithm with high computational efficiency, it is sensitive to the distribution of the pilot. Small deviations from the independent and identically distributed (i.i.d.) sub-Gaussian random variable model of the pilot can cause the algorithm to diverge. To address this issue, reference \cite{8713501} developed a vector AMP algorithm (VAMP) and proved that VAMP has strict scalar state evolution, which is applicable to a wider class of pilot distributions. 
\item \textit{SBL}: SBL is a popular approach for single sparse signal recovery from a Bayesian perspective, and has empirically shown superior recovery performance \cite{tipping2001sparse,1315936}. Gaussian generalized approximate message passing (GGAMP) is incorporated into SBL in \cite{8074806}, and the approach is further extended from the single measurement vector (SMV) to the multiple measurement vectors (MMV). Numerical experiments show that the proposed algorithm has great robustness and computational advantages. Moreover, in \cite{8039502} considered a general joint-sparse model and provided a Bayesian approach that extends the SBL for the joint-sparse model.

\item \textit{Other CS-based methods}: Taking advantage of the total squared coherence of pilot sequences and convex relaxation, a novel pilot design approach was proposed for joint AUD and CE in \cite{8269096}. The proposed design exhibits superior performance in relation to the random pilot design and the design based on tight frames \cite{6457477}. According to CS theory, the overhead of CS-based methods highly depends on the sparsity and the number of devices. A more sparse representation of the active device’s channel can be obtained by using dictionary learning. In view of the distinct wireless propagation environments in different cells, a dictionary-learning-enhanced AUD and CE method for massive MIMO systems was proposed in \cite{9839006}, which uses a dictionary learned from the historical channel information of the specific cell. A dictionary learning-based uncoordinated access method was proposed to jointly identify the set of active devices and detect data symbols embedded in signals transmitted by active devices in \cite{8903537}. 
\end{itemize}

\textbf{CS-Based Methods Using Additional Information:}
Additional information has been used for AUD and CE to achieve better performance. Various additional information could be
exploited, including constellation information, active user temporal correlation, channel spatial correlation, and so on.
\begin{itemize}
\item \textit{Use prior information:} Devices might have different service requirements, which in turn leads to different data lengths for transmission. The data length diversity was exploited in \cite{9268113} to enhance AUD and CE. In \cite{9267798}, different prior information was used simultaneously to improve the performance of MRA. In specific, the joint sparsity between the channel and the data was used to improve the accuracy of pilot detection, and the modulation information and the cyclic redundancy check were used for channel correction to improve the performance of
data recovery. Reference \cite{8761393} improved the performance of AUD and CE using side information (SI) brought from data recovery, and improved the performance of data recovery in SIC using the
SI brought from AUD.
\item \textit{Use temporal correlation:} Modified subspace pursuit (M-SP) was proposed to exploit a prior support and the prior support quality information in massive MIMO CE \cite{rao2015compressive}. Instead of blindly using the prior support, the proposed approach utilizes it adaptively by taking the quality information into account. Taking advantage of the temporal correlation of the channel, the required pilot overhead was further reduced. Moreover, they proved that an appropriate prior support can lead to better CS recovery performance and further propose some robustness designs to combat incorrect prior support quality information. The channel response estimated in the previous slots was used as SI when estimating the channel response for the current slots in \cite{ma2019approximate}. An algorithm called AMP-SI was proposed to embed the SI into the denoiser design under the SMV-AMP framework for single-antenna systems. In \cite{9892680}, it was proposed to leverage the SI of the temporal correlation in the activity of the device, that is, the explicit quantification of the correlation between the activity mode estimated by the AMP algorithm in the former coherent block and the actual activity mode in the current coherent block. The SI was applied to the AMP denoiser design, and a new SI-aided AMP framework was derived. Numerical results showed that the proposed
method can significantly improve the performance of AUD and CE.
\item \textit{Use channel spatial correlation:} An alternating direction multiplier method (ADMM) using prior information on second-order statistics, that is, spatial channel covariance of each user channel, was proposed to improve performance in \cite{djelouat2021joint}. In specific, the original problem of AUD and CE can be formulated as:
\begin{flalign}
\min _{\mathbf{X}} \frac{1}{2}\left\|\mathbf{S} \mathbf{X}-\mathbf{Y}\right\|_{\mathrm{F}}^{2}+\beta_{1}\left\|\mathbf{X}\right\|_{2,1},
\end{flalign}
where $\mathbf{Y}$, $\mathbf{S}$ and $\mathbf{X}$ denote the received signal matrix, the pilot matrix/sensing matrix and the effective channel matrix, respectively. $\|\cdot\|_{F}$ represents the Frobenius norm. $\|\cdot\|_{2,1}$ is a matrix norm that sums up the $\ell_{2}$ norm of all rows in the matrix. $\beta_{1}$ denotes a trade-off between the fidelity term and the row-sparsity level of $\mathbf{X}$. By using the information embedded in the estimated channel covariance matrix, the problem can be reformulated as:
\begin{align}
&\min _{\mathbf{X}} \frac{1}{2}\left\|\mathbf{S} \mathbf{X}-\mathbf{Y}\right\|_{\mathrm{F}}^{2}+\beta_{1}\|\mathbf{X}\|_{2,1}\notag\ \\
&+\beta_{2} \sum_{i=1}^{N} \mathbb{I}\left(\mathbf{x}_{i}\right)\left\|\mathbf{x}_{i} \mathbf{x}_{i}^{\mathrm{H}}-\tilde{\mathbf{R}}_{i}\right\|_{\mathrm{F}}^{2},\label{prob-use-spatial-correlation}
\end{align}
where $\tilde{\mathbf{R}}_{i}$ is the estimated scaled covariance matrix, $\mathbf{x}_{i} \mathbf{x}_{i}^{\mathrm{H}}$ is the sample covariance matrix, and $\beta_{1}$ and $\beta_{2}$ are parameters that control the trade-off. $\mathbb{I}\left(\mathbf{x}_{i}\right)$ is defined as follows:
\begin{flalign}
\mathbb{I}\left(\mathbf{x}_{i}\right)=\left\{\begin{array}{l}
1, \quad\left\|\mathbf{x}_{i}\right\|_{2}>0, \\
0, \quad\left\|\mathbf{x}_{i}\right\|_{2}=0,
\end{array} \quad \forall i \in \mathcal{N}\right.
\end{flalign}
where $\mathcal{N}=\{1, \ldots, N\}$. The extra channel space correlation in (\ref{prob-use-spatial-correlation}) improves the AUD and CE performance with lower signaling overhead.
\item \textit{Using constellation information:} Constellation information has been used to improve AUD and CE performance in literature. In \cite{di2019activity}, a low complexity detection algorithm for MMTC is proposed, namely, activity-aware low complexity multiple feedback successive interference cancelation (AA-MF-SIC). The main idea of this method is to evaluate the reliability of each estimated symbol. If the current estimate fails to satisfy some given constraint, other possible constellation points are then considered with the cost of increased computational complexity. In \cite{9268113}, backward sparsity adaptive matching pursuit with checking and projecting (BSAMP-CP) was proposed to jointly carry out sparsity level estimation, AUD, CE and data recovery in two phases. Specifically, in the first phase, it conducted the sparsity level estimation in a backward manner, exploiting the data length diversity of active users. In the second phase, it jointly conducted AUD, CE and data recovery, taking the joint sparsity information of pilot and data symbols, the error checking information and the modulation constellation information into account. Using the prior information, the problem formulated in \cite{9268113} can be expressed as:
\begin{subequations}
\begin{align}
\min _{\mathbf{H}, \mathbf{D}} & \|\mathbf{Y}-\mathbf{S} \mathbf{X}\|_{\mathrm{F}} \label{14a}\\
\text { s.t. } & \|\mathbf{X}\|_{\text {row }, 0} \leq K ,\label{14b}\\
& \mathbf{X}=\mathbf{H}[\mathbf{P}, \mathbf{D}] , \label{14c}\\
& \forall k=1 \ldots N ,\label{14d}\\
& \mathrm{g}\left(\mathbf{D}_{(\{k\},:)}\right)=1 , \label{14e}\\
& \mathbf{D}_{(\{k\},:)} \in(\mathcal{A} \cup \mathcal{O})^{n_{d}} ,\label{14f}
\end{align}
\end{subequations}
where $\|\cdot\|_{\text {row }, 0}$ represents $\ell_{\text {row }, 0}$-semi norm that outputs the number of non-zero rows of the input matrix, and $K$ denotes the number of active users. $\mathbf{H}$ denotes the channel coefficients of all devices, and $\mathbf{P}$ and $\mathbf{D}$ denote the pilot matrix and data matrix of device $k$, respectively. The constraints (\ref{14b}) and (\ref{14c}) describe the joint sparse characteristics of pilot and data.  The constraint (\ref{14e}) describes the checking mechanism when data recovery is implemented, where
the function $\mathrm{g}(\cdot)$ denotes the error checking procedure. The constraint (\ref{14f}) uses constellation point information, where $\mathcal{A}$, $\mathcal{O}$ and $n_d$ denote the transmitted symbol set, the set of zeros and the maximum number of transmitted data symbols, respectively. By exploiting backward activity level estimation that exploits data length diversity, this method achieved superior performance.
\end{itemize}

\textbf{Methods Using Structured Access Protocols:}
Reference \cite{paolini2015coded} exploited coded slotted aloha (CSA) and proposed the paradigm of coded random access, in which the structure of the access protocol can be mapped to a structure of an erasure-correcting code defined on a graph. This provided the possibility of designing efficient random access protocols using coding theory. A novel grant-free NOMA method called CTSMA was proposed in \cite{ma2018coded}. This method divides each user’s data packet into multiple segments for processing, so that conflicts occur only on segments using the same extension code, which can be resolved with redundant segments. For the MMTC scenario in industrial IoT, a multislot design scheme was proposed with joint CTSMA and CSA to further promote collision resolution capability \cite{ma2018joint}. Performance analysis has shown the superiority of multislot design in the short uplink contention period. A coded CS (CCS) was proposed in \cite{amalladinne2020coded}, in which each active device divides its data into several subblocks and then adds redundancy using system linear block codes. CS was used to restore the order of the subblocks and splice them together to obtain the original information. A pilot-based coherent scheme was proposed in \cite{fengler2022pilot} for unsourced random access in the MIMO scenario, where the user messages are divided into two parts. Specifically, several bits are used to select a shorter code word from the codebook, and then encode the remaining message bits. The receiver uses the MMV-AMP algorithm to estimate the channel from the pilot part, and then decodes the second part with the maximum ratio combining (MRC). A signal scrambling based joint blind channel estimation, activity detection, and data decoding (SS-JCAD)
scheme is proposed for coded massive random access in \cite{10439027}. Signal scrambling technique is used and treated as the user-specific signature. A simple yet efficient receiver is designed which integrates the blind CSI estimation module with the forward error correction decode to achieve joint blind channel estimation, activity detection, and data decoding. The authors of \cite{10473114} propose a high-efficiency massive uncoupled unsourced random access scheme for 6G wireless networks without requiring extra parity check bits. A low-complexity Bayesian joint decoding algorithm was designed to implement codeword detection and stitching based on channel statistical information.

A summary of existing work on sparse recovery based methods for MRA is given in Table \ref{SRBM4RA}.
\renewcommand\arraystretch{1.2}
\begin{table*}[htp]
\begin{center}
\caption{Summary of Existing Work on Sparse Recovery Based Methods for MRA}
\label{SRBM4RA}
\begin{tabular}{|m{3.1cm}<{\centering}|m{2.5cm}<{\centering}|m{9.6cm}|m{1.2cm}<{\centering}|}
\hline
\textbf{Category} & \textbf{Algorithm} & \textbf{Characteristics}                                                                   & \textbf{Ref.}    \\ \hline
\multirow{11}{*}{CS-based typical methods} & ISTA,FISTA         & Classic method for solving LIP                  & \cite{daubechies2004iterative,WOS:000278101000009} \\ \cline{2-4} 
& SBL  & Provide an efficient algorithm that extends SBL for a joint-sparse model                       & \cite{8039502}         \\ \cline{2-4} 
& OMP  & Using OMP algorithm for sparse signal recovery                                             & \cite{tropp2007signal}         \\ \cline{2-4}
& OMP  & Use OMP and orthogonal least squares (OLS) for MUD                                             & \cite{schepker2011sparse}         \\ \cline{2-4}
& GOMP  & Use GOMP to deal with asynchronous case                                                       & \cite{bockelmann2013compressive}         \\ \cline{2-4}
& OMP  & Propose a contention-based MRA scheme to support massive access                                & \cite{9148341}         \\ \cline{2-4} 
& AMP  & Use AMP algorithm for AUD and CE and provides analytical characterization                      & \cite{8264818}         \\ \cline{2-4} 
& VAMP & VAMP has strict state evolution and stronger robustness compared with AMP                      & \cite{8713501}         \\ \cline{2-4}
& Pilot design algorithm  & Propose a novel pilot design approach for joint AUD and CE                      & \cite{8269096}         \\ \cline{2-4} 
& Dictionary learning  & Use dictionary learning to enhance AUD and CE in massive MIMO system                      & \cite{9839006}         \\ \cline{2-4} 
& Dictionary learning  & Propose dictionary learning-based uncoordinated access method to reduce the pilot overhead & \cite{8903537}         \\ \hline
\multirow{8}{*}{\makecell[l]{CS-based methods of using\\additional information}} 
& M-OMP   & Use various prior information to enhance performance                                        & \cite{9267798}         \\ \cline{2-4} 
& FSP-RIC & Use SI brought from the data recovery                                                       & \cite{8761393}         \\ \cline{2-4}
& M-SP   & Use channel time correlation to reduce pilot overhead                                        & \cite{rao2015compressive}         \\ \cline{2-4} 
& AMP-SI & Use channel response in the previous slots as SI to estimate the cerrent channel response    & \cite{ma2019approximate}         \\ \cline{2-4}
& AMP    & Use the temporal correlation in device activity as SI and apply SI into AMP denoiser design  & \cite{9892680}             \\ \cline{2-4} 
& ADMM   & Utilize prior information on the second-order statistics to improve performance              & \cite{djelouat2021joint}    \\ \cline{2-4} 
& AA-MF-SIC   & Use constellation point information to evaluate the reliability of the estimation       & \cite{di2019activity}    \\ \cline{2-4} 
& CSAMP-CP   & Divide AUD, CE and data recovery into two phases                                         & \cite{9268113}         \\  \hline 
\multirow{7}{*}{\makecell[l]{Methods using structured\\access protocols}} 
& CSA                & Propose the paradigm of coded random access                                      &\cite{paolini2015coded}         \\ \cline{2-4}                              
& CTSMA  & Divide data packet into multiple segments for processing                                     & \cite{ma2018coded}\\ \cline{2-4}
& CTSMA  & Propose a scheme joint CTSMA and CSA to promote the collision resolution capability & \cite{ma2018joint}\\ \cline{2-4}
& CCS  & Divide the data into subblocks and use algorithms to recover and splice it & \cite{amalladinne2020coded}  \\ \cline{2-4}
& MMV-AMP  & Split the user messages into two parts and use MMV-AMP and MRC for decoding  & \cite{fengler2022pilot}  \\
\cline{2-4}
& SS-JCAD  & Use the scrambling pattern as a user-specific signature  & \cite{10439027}  \\
\cline{2-4}
& OAMP  & Implement codeword detection in the case of unknown priors and codeword stitching with the assistance of channel statistics  & \cite{10473114}  \\
\hline
\end{tabular}
\end{center}
\end{table*}
\renewcommand\arraystretch{1}

\subsubsection{Structured Sparse Recovery-Based Methods for MRA}
Unlike traditional sparse recovery methods that focus solely on identifying individual sparse elements, structured sparse recovery goes a step further by exploiting the inherent structures or patterns presented in the signal. Structured sparse recovery-based methods represent a compelling alternative that harnesses the power of structures to unlock deeper insights and achieve superior performance for MRA. Moreover, it enhances the efficiency of the recovery algorithm, reduces computational complexity, and enables scalable solutions for MRA problems.

\textbf{Methods Using Row Sparsity:}
For the BS equipped with a single antenna, the joint AUD and CE can be expressed as an SMV problem in CS. In the case of the BS with multiple antennas, the joint AUD and CE turns into an MMV problem.
The user activity pattern is the same for all signals received by different BS antennas, which results in a row sparsity structure in the joint AUD and CE problem. In \cite{8264818,8323218,8320821}, AMP algorithms were applied to the multi-antenna scenario. AMP with vector denoiser and parallel AMP-MMV were used for joint AUD and CE in \cite{8264818}, with simulation results that demonstrate a significant performance improvement caused by the deployment of multiple antennas in BS. According to the theoretical analysis in \cite{8323218,8320821}, when the number of BS antennas tends to infinity, perfect AUD can be achieved, which indicates that the sparse signal recovery-based method is suitable for the MRA problem. Reference \cite{7248738} considered joint AUD and CE in uplink cloud radio access network (CRAN) systems with few active users and proposed a modified Bayesian compressive sensing algorithm (BCS), which exploits not only the sparsity of the active users, but also the prior channel information of path loss effects and the joint sparsity structures. Moreover, in \cite{9174826}, Cheng et al. proposed an MMV-based orthogonal AMP (OAMP), and derived the group Gram-Schmidt orthogonalization  (GGSO) procedure to implement OAMP. It has been shown that OAMP is superior to AMP when pilot sequences are generated using the Hadamard pilot matrix. Most previous work considered narrow-band MRA scenarios assuming single-carrier transmission. Reference \cite{ke2020compressive} investigated grant-free MRA in massive MIMO-OFDM systems and proposed a Turbo-GMMV-AMP algorithm running in an alternating manner for joint AUD and CE, which jointly exploits the space-frequency-angular-domain sparsity of the MRA channel matrix to enhance performance. In \cite{9566698}, Chen et al. proposed a contention-based joint AUD and CE method for massive MIMO systems with angular domain enhancement. The problems of AUD and CE were presented as LIP with simultaneous row sparse and cluster sparse structures, in which angle of arrival (AOA) information was used to enhance AUD and CE. They used SBL to formulate the optimization problem, which is given as:
\begin{equation}\label{structured-MRA-SBL}
\begin{split}
&\min _{\mathbf{x}}\|\mathbf{y}-\mathbf{S} \mathbf{x}\|_{2}^{2}+\lambda f(\mathbf{x}) \\
&\text { s.t. } f(\mathbf{x})=\min _{\boldsymbol{\omega} \geq 0} \mathbf{x}^{T} \boldsymbol{\Omega} \mathbf{x}+\log |\boldsymbol{\Sigma}| ,
\end{split}
\end{equation}
where $\lambda$ denotes the variance of Gaussian noise, $\boldsymbol{\Omega} \triangleq \operatorname{diag}\left(\omega_{1}, \ldots, \omega_{N}\right) \in \mathbb{R}^{N \times N}$, $\boldsymbol{\omega}$ is a vector of hyperparameters governing the prior variance of the elements in signal and $\boldsymbol{\Sigma}=\lambda \mathbf{I}+\mathbf{S}\boldsymbol{\Omega} \mathbf{S}^{T}$. By solving the bilevel optimization problem in (\ref{structured-MRA-SBL}), multiple active users can be detected successfully, together with accurate CE, even when they use the same time-frequency-code resource for random access. For bilevel optimization, interested readers may refer to \cite{10502023}.

\textbf{Methods Using Block/Group Sparsity:} 
Block/group sparsity generalizes the standard sparse signal model by separating the elements into a number of blocks/groups. By combining CS based multi-user detection (CS-MUD) with multi-carrier access schemes, a new solution was proposed to leverage the group sparsity structure through the GOMP \cite{7145755}. In \cite{8482464}, the joint AUD and CE in grant-free NOMA systems was formulated as a block CS-based sparse signal recovery problem. The authors made explicit use of the block sparsity inherent in the equivalent block-sparse model and proposed two enhanced block CS-based greedy algorithms, namely, threshold aided block sparsity adaptive subspace pursuit (TA-BSASP) and cross-validation aided block sparsity adaptive subspace pursuit (CVA-BSASP). In \cite{8419284}, the authors proposed block orthogonal matching pursuit (BOMP) algorithm for AUD and CE. The algorithm transforms the original model, and then exploit the block sparsity of matrix $\mathbf{X}$. Based on the maximal Euclidean norm, the objective function in the $l$-th iteration of the algorithm was given as:
\begin{flalign}
\mathcal{I}_{l}=\underset{\mathcal{I}_{B} \in \mathcal{I}_{\text {Block }}}{\operatorname{argmax}}\left\|\overline{\mathbf{P}}_{\left[\mathcal{I}_{B}\right]}^{H} \mathbf{r}_{l-1}\right\|_{2},
\end{flalign}
where $\mathcal{I}_{l}$ is the selected index set, $\mathcal{I}_{\text {Block }}$ is the union of all block indices, $\overline{\mathbf{P}}$ is obtained from $\mathbf{X}$ through a series of transformations and $\mathbf{r}_{l-1}$ is the residue of the $(l-1)$-th iteration. Furthermore, a block sparse Bayesian learning (BSBL) algorithm was proposed, which outperformed BOMP in terms of computational complexity and recovery performance. As AMP is sensitive to the choice of pilot sequence while having low complexity, AMP-BSBL algorithm was proposed in \cite{10042415}, which retains the advantages of AMP and is robust to the choice of pilot sequence. In \cite{10064098,10349818}, a goal-oriented detection approach is proposed to exploit the continuous angular group sparsity feature of massive MIMO wireless channels. For stationary users with known angular information, their activity can be identified according to the active angular signals without reconstructing the corresponding channels and messages. 

\textbf{Methods Using Other Structures:}
There are methods that use other structures to reduce the degree of freedom in the problem. A low-complexity multiuser detector based on structured CS, namely, the structured iterative support detection (SISD) algorithm, utilized the inherent structural sparsity of user activities in a NOMA system to realize joint AUD and CE \cite{7462187}. Simultaneous SBL for joint sparse approximation with two structured sparse models (SSMs) were proposed in \cite{7558157}, where one is row-sparse with embedded element-sparse and the other one is row-sparse plus element-sparse, which can be extended for AUD and CE in some special cases. Taking advantage of a more accurate prior distribution to characterize the structured sparsity of access signals, efficient joint activity and data detection algorithms, namely, the OAMP MMV algorithm with simplified structure learning and accurate structure learning, were proposed in \cite{mei2021compressive}. Furthermore, it can be extended to SIC-based detection with channel coding to achieve highly reliable random access. An expectation-maximization-aided generalized AMP algorithm with a Markov random field support structure, namely, EM-MRF-GAMP, was proposed in \cite{9721077}, which captures the inherent clustered sparsity structure of the angular domain channel. Then, message reconstruction in the form of a clustering decoder was performed by recognizing slot-distributed channels of each active user based on similarity. The proposed scheme achieved better error performance compared to traditional CS-based random access schemes.

\subsubsection{Low-Rank Matrix/Tensor Recovery-Based Methods for MRA}
Low-rank matrix/tensor recovery-based methods, leveraging the inherent structure and sparsity present in the channel, offer a highly effective way to address the challenges in MRA. In \cite{haghighatshoar2018improved,chen2021phase,chen2019covariance,fengler2021non,chen2021sparse}, by exploiting the low rank and statistical characteristics of the signal, covariance-based methods for grant-free MRA were used in AUD. The sparse activity detection problem was formulated as a large scale fading estimation problem. Let $a_{n} \in\{1,0\}$ denote the activity of device $n$, $\mathbf{h}_{n}$ denote the channel vector between device $n$ and the BS, where $\mathbf{h}_{n} \in \mathbb{C}^{M \times 1}$ include both Rayleigh fading component and large scale fading component, $M$ denotes the number of antennas and $L$ is the pilot length. The received signal $\mathbf{Y} \in \mathbb{C}^{L \times M}$ at the BS can be expressed as
\begin{flalign}
\mathbf{Y}=\sum_{n=1}^{N} a_{n} \mathbf{s}_{n} \mathbf{h}_{n}^{T}+\mathbf{Z} \triangleq \mathbf{S} \boldsymbol{\Gamma}^{\frac{1}{2}}\mathbf{H}+\mathbf{Z},
\end{flalign}
where $\mathbf{S} \triangleq\left[\mathbf{s}_{1}, \mathbf{s}_{2}, \cdots, \mathbf{s}_{N}\right] \in \mathbb{C}^{L \times N}$, $\boldsymbol{\Gamma} \triangleq \operatorname{diag}\left(\gamma_{1}, \ldots, \gamma_{N}\right) \in \mathbb{R}^{N \times N}$ with $\gamma_{n}=\left(a_{n} g_{n}\right)^{2}$ is a diagonal matrix and $\mathbf{H} \triangleq\left[\mathbf{h}_{1} / \sqrt{g_{1}}, \mathbf{h}_{2} / \sqrt{g_{2}}, \cdots, \mathbf{h}_{N} / \sqrt{g_{N}}\right]^{T} \in \mathbb{C}^{N \times M}$ is the normalized channel matrix with $g_n$ is the large scale fading component. The user activity detection can be formulated as a maximum likelihood estimation (MLE) problem. After normalization and simplification, the objective function can be written as follow:
\begin{equation}
\begin{split}
\underset{\boldsymbol{\gamma}}{\text{min}} &\ \log |\boldsymbol{\Sigma}|+\operatorname{tr}\left(\boldsymbol{\Sigma}^{-1} \hat{\boldsymbol{\Sigma}}\right) \\
\text {s.t. } &\ \boldsymbol{\gamma} \geq 0,
\end{split}
\end{equation}
where $\hat{\boldsymbol{\Sigma}} \triangleq \frac{1}{M} \mathbf{Y} \mathbf{Y}^{H}=\frac{1}{M} \sum_{m=1}^{M} \mathbf{y}_{m} \mathbf{y}_{m}^{H}$
is the average sample covariance matrix of the received signal on different antennas, $\boldsymbol{\Sigma}=\mathbf{S} \boldsymbol{\Gamma} \mathbf{S}^{H}+\sigma_{w}^{2} \mathbf{I}$ is the covariance matrix,  $\boldsymbol{\gamma} = \left[\gamma_{1}, \ldots, \gamma_{N}\right]^{T}$ is the indicator of activity and large-scale fading of all devices. The coordinate descent method can be used to find good solutions of this optimization problem. An effective active set algorithm was proposed following reference \cite{chen2019covariance,chen2021phase} to reduce the computational complexity in \cite{wang2021efficient}. Furthermore, in \cite{chen2021phase,chen2019covariance} it was shown that covariance-based methods can significantly shorten the minimum pilot sequence length for AUD when CE is not required. In \cite{decurninge2020tensor,decurninge2021tensor}, tensor-based modulation (TBM) was introduced to unsourced random access, where data decoding exploits low-rank tensor structure and tensor decomposition. A dimension reduction method was proposed to use the sparse and low-rank structure to project the original device state matrix into a low-dimension space in \cite{shao2019dimension}. Then an optimized design framework with coupled full column rank constraints was developed to reduce the size of the search space, leading to a shorter pilot sequence than the original sparse recovery-based methods.

A summary of existing work on structured sparse recovery-based methods for MRA is given in Table \ref{SSRBM4RA}.

\begin{table*}[htp]
\renewcommand\arraystretch{1.2}
\begin{center}
\caption{Summary of Existing Work on Structured Sparse Recovery-Based Methods for MRA}
\label{SSRBM4RA}
\begin{tabular}{|m{2.5cm}<{\centering}|m{1.6cm}<{\centering}|m{10.05cm}|m{1.35cm}<{\centering}|}
\hline
\textbf{Category} & \textbf{Algorithm} & \textbf{Characteristics} & \textbf{Ref.}    \\ \hline
\multirow{6}{*}{\makecell[l]{Methods using the\\row sparsity}} & AMP & Employ two different AMP algorithms, namely, the AMP with vector denoiser and the parallel AMP-MMV      & \cite{8264818}         \\ \cline{2-4} 
& AMP                & Use AMP algorithm to solve MMV problem, joint AUD, CE and data transmission & \cite{8323218,8320821} \\ \cline{2-4} 
& BCS                & Joint sparsity structures in multi-antenna uplink CRAN systems              & \cite{7248738}         \\ \cline{2-4} 
& OAMP               & Propose the OAMP-MMV form, derive GGSO procedure to implement OAMP          & \cite{9174826}         \\ \cline{2-4} 
& GMMV-AMP           & Use angular domain sparsity to enhanced performance in MIMO-OFDM system     & \cite{ke2020compressive}         \\ \cline{2-4} 
& SBL                & Use simultaneous row sparse and cluster sparse structures to enhance AUD and CE in massive MIMO system & \cite{9566698}         \\ \hline
\multirow{5}{*}{\makecell[l]{Methods using the\\block/group sparsity}} 
& GOMP               & Combine CS-MUD with multi-carrier access schemes                            & \cite{7145755}         \\ \cline{2-4} 
& BSP                & Exploit the block sparsity and propose two algorithms, namely, TA-BSASP and CVA-BSASP    & \cite{8482464}         \\ \cline{2-4} 
& MP-BSBL            & Combine BSBL with message passing and easily used to data detection         & \cite{8419284}         \\ \cline{2-4} 
& AMP-BSBL           & Retain the advantages of BSBL and has low complexity                       & \cite{10042415}         \\ \cline{2-4} 
& ADMM and Clustering               & Utilize wireless channels' angular continuous group-sparsity feature and knowledge of stationary users' angular information              & \cite{10064098}         \\ \hline
\multirow{4}{*}{\makecell[l]{Methods using other\\structures}}                         & SISD               & Exploit the inherent structured sparsity of user activity naturally existing in NOMA systems        & \cite{7462187}         \\ \cline{2-4} 
& SBL                & Propose two SSMs based on SBL                                               & \cite{7558157}         \\ \cline{2-4} 
& OAMP-MMV           & Exploit a more accurate prior distribution to characterize the structured sparsity of signals   & \cite{mei2021compressive}         \\ \cline{2-4} 
& EM-MRF-GAMP        & Exploit the inherent clustered sparsity structure in the angular domain in MIMO system          & \cite{9721077}         \\ \hline
\end{tabular}
\end{center}
\end{table*}
\renewcommand\arraystretch{1}

\subsection{Machine Learning Empowered MRA}

\begin{figure}[!t]%
\centering%
\includegraphics[width=0.4\textwidth]{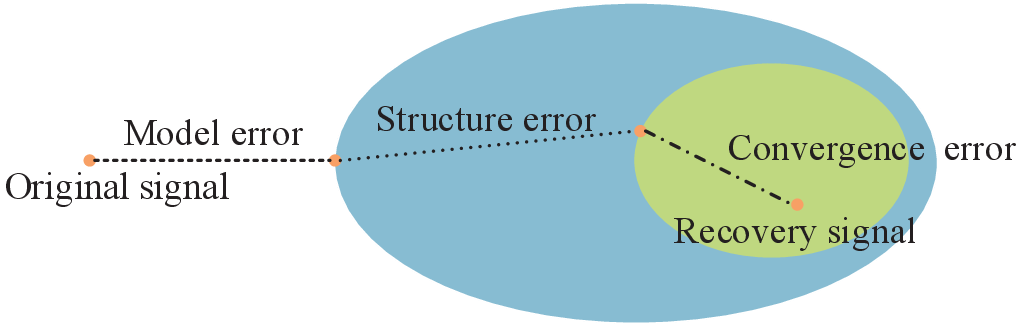}%
\DeclareGraphicsExtensions. \caption{The decomposition of the error in model-driven approaches \cite{BAI2020107729}. }
\label{fig:error}
\end{figure}%
\par
Deep learning (DL) is a promising approach to solving complex optimization problems. In particular, by unfolding an iterative algorithm into a neural network (NN), we can learn the parameters of iterative algorithms from training data, which differs from traditional algorithms that employ predetermined parameters. This approach is also called deep unfolding. Using DL for MRA has several advantages. One prominent advantage is the increase in the speed of convergence. A fast DL-based approach for CS-based multi-user detection in massive MTC systems shows more than a 10-fold decrease in computing time, compared to traditional iterative algorithms \cite{8891371}. Reducing the computing time is critically important for latency sensitive applications in 5G and 6G. In addition, appropriately trained DL networks are capable of increasing the detection/estimation accuracy. As shown in Fig.~\ref{fig:error}, the detection/estimation error of the traditional model-driven approach results is composed of model error, structure error and convergence error in general. The model error comes from imperfect modeling of the original engineering problem, i.e., oversimplification of the real problem. The structure error comes from approximation of the original objective function, e.g., using convex relaxation to facilitate algorithm derivation for complex problems. The optimal solution of the approximated problem would deviate from the true ``optimal'' one of the original problem. The convergence error comes from iterative algorithms that fall into local optima in the function landscape. DL-based methods learn the mapping from the input to the output directly and has the potential to relieve challenges brought by the model error, the structure error and the convergence error in model-driven approaches. Reference \cite{BAI2020107729} provided a comprehensive review of recent progress in the development of DL for solving linear inverse problems.

\par
Significant efforts have been put into designing a powerful receiver with the assistance of DL in the past few years. A fast DL-based approach for AUD in MRA was first proposed in \cite{8891371}, where a nonlinear block restrictive activation unit was proposed to capture the block sparse structure in wide-band wireless communication systems (or multi-antenna systems). Reference \cite{9374107} proposed to aggregate two Zadoff-Chu preamble sequences from two different roots to obtain a larger set of preambles and used DL to decode the aggregate preambles. DL-based activity detection for cell-free (CF) massive multiple input multiple output (mMIMO) systems was investigated in \cite{10143353}, and a transfer learning-based ensemble model was established in the central processing unit to achieve a better global detection decision. Various DL-based methods \cite{9432908,9634107,9484069,9508782,9863071,9791132} have been developed for joint AUD and CE have been developed. Reference \cite{9605579} introduced an adaptive tuning module and designs a feature-assisted adaptive tuning DL method that includes inner and outer networks to solve the massive device detection problem. In \cite{9484069,9791132}, a deep unfolding method based on the traditional ADMM algorithm was developed, where the iteration operation was unfolded into the network layer. Reference \cite{9791132} further considered a model-driven neural network architecture based on vector AMP. The long short-term memory (LSTM) was exploited for the detection of active devices and the estimation of the channel of active devices in \cite{9634107}. In \cite{9508782}, the joint AUD and CE problem was formulated as a group-sparse matrix estimation problem, and a simplified unrolled recurrent neural network with reduced training parameters was proposed to solve the problem. Correlated antennas at the BS was considered in the design of the DL method for MRA in \cite{9863071}. Joint activity detection, CE, and data recovery were investigated in \cite{9685696,9605579}. A novel dual network was proposed in \cite{9685696} to utilize the sparse consistency between the channel vector and the data matrix of all active users. Information distilled from the initial data recovery phase was employed as prior information to improve DL methods for activity detection and CE, which in turn improves data recovery performance. Theoretical analysis demonstrated that the proposed method can converge faster and support more users \cite{9605579}.

\par
Pilot sequences can be jointly designed in DL-based methods for MRA. In \cite{9174792}, two DL models, using the standard autoencoder structure in DL, were investigated for the joint design of pilot sequences and CE methods, and the joint design of pilot sequences and device activity detection in MIMO-based grant-free random access, respectively. For small packet transmission, CE becomes costly and inevitably introduces a huge amount of overhead. Several existing works have attempted to investigate non-coherent schemes, which do not need CSI at the receiver. In \cite{ma2023model}, the transmitted messages were embedded in the index of the transmitted pilot sequence of each active device, and a DL-modified AMP network was developed to effectively exploit the correlation of pilot activity. In \cite{9682613}, inspired by the algorithm based on AMP, a model-driven DL method was proposed for noncoherent communications to jointly detect user activity and desired data.

\par
DL-based methods have also been investigated for the MRA problem with imperfect synchronization. In \cite{9390399}, a learned AMP network was designed for joint user activity detection, delay detection and CE. Asynchronous access of machine-type devices that send data packets of different frame sizes, called data length diversity, was considered in \cite{10304065}. The composite problem of AUD, CE, and data recovery was formulated as a structured sparse recovery problem, and a DL method unfolding from AMP with a backward propagation algorithm (AMP-BP) was developed to this end.

\par
The above DL-based methods for MRA consider a fixed number of layers. However, they ignore a key character in traditional iterative algorithms, where the number of iterations required for convergence changes with varying sparsity levels, i.e.,  ``easy'' tasks should be solved with fewer iterations. In MRA, the number of active users is difficult to predict and could vary in a large range depending on service type and user mobility. For a relatively small number of active users, using a fixed number of iterations leads to the waste of computing power and increases communication latency. For a relatively large number of active users, using a fixed number of iterations leads to poor user detection accuracy, as the algorithm is not likely to converge yet. To this end, a novel depth adaptive approach was developed to effectively reduce the computational complexity of DL-based methods \cite{9252937}. The proposed method with adaptive depth provided the solution to adapt the neural network to the varying conditions in the number of access users.

\par
Another line of research to improve MRA efficiency is the tuning of access configurations, which controls the probability of user access and alleviates access contention. DL-based transmit power control schemes were proposed in \cite{8472885,8428396}. Unlike conventional transmit power control schemes in which complex optimization problems have to be solved in an iterative manner, the DL-based scheme determines the transmit power with a relatively low computation time. The joint design of the transmit power control functions and the estimation algorithm was proposed in \cite{9590505}, which improves the accuracy of AUD and CE. A DL-based scheme was proposed to detect and resolve preamble collisions in \cite{9449922}. Given the detection result, the preamble collision can be resolved according to the different timing advance, which is also realized by DL. In \cite{9900130}, DL was employed to detect random access collisions by learning the features of the signals received in the BS and then an additional contention process was scheduled for users with collisions. A summary of existing work on machine learning for MRA is given in Table \ref{Summary-ML-RA}.

\begin{table*}[htb]   
\renewcommand\arraystretch{1.2}
\begin{center}   
\caption{Summary of Existing Work on Machine Learning for MRA.}  
\label{Summary-ML-RA} 
\begin{tabular}
{|m{3.5cm}<{\centering}|m{11cm}|m{1.9cm}<{\centering}|}
\hline
\textbf{Category}   & \textbf{Characteristics}  & \textbf{Ref.}     \\ \hline
\multirow{2}{*}{AUD}  & Exploit block restrictive activation nonlinear unit to capture the block sparse structure & \cite{8891371} \\ \cline{2-3} 
 & Establish a transfer learning-based ensemble model in CF mMIMO system   & \cite{10143353}   \\ \hline
\multirow{4}{*}{Joint AUD and CE} & Exploit LSTM   & \cite{9634107}          \\ \cline{2-3} 
 & Use deep unfolding method based on ADMM   & \cite{9484069,9791132} \\ \cline{2-3} 
  & Exploit group-sparse property to simplify network    & \cite{9508782}  \\ \cline{2-3} 
 & Employ initial data recovery phase as prior infomation & \cite{9685696}  \\ \hline
\multirow{3}{*}{Joint AUD and pilot design}  & Use standard auto-encoder structure & \cite{9174792} \\ \cline{2-3} 
& Develop a DL-modified AMP network to exploit the pilot activtity correlation  & \cite{ma2023model}         \\ \cline{2-3} 
 & Propose GAMP-inspired DL method   & \cite{9682613}  \\ \hline
\multirow{2}{*}{Imperfect synchronization}  & Design a learned AMP network for joint user activity detection, delay detection, and CE & \cite{9390399} \\ \cline{2-3} 
 & Consider data length diversity and formulate a structured sparse recovery   & \cite{10304065}   \\ \hline
Unknown sparsity level                                        & Develop a novel depth adaptive approach to adapt different number of access users                       & \cite{9252937}          \\ \hline
\multirow{3}{*}{Access configuration}   & Use DL scheme to replace conventional power control schemes & \cite{8472885,8428396} \\ \cline{2-3} 
& Jointly design the transmit power control functions and the estimation algorithm   & \cite{9590505}    \\ \cline{2-3} 
& Resolve random access collision by DL & \cite{9449922,9900130} \\ \hline
\end{tabular}
\end{center}   
\end{table*}
\renewcommand\arraystretch{1}

\section{Non-Orthogonal Multiple Access}
In this section, we introduce progress in model-based methods and learning empowered methods for NOMA. 

\subsection{Model-Based Methods for NOMA}
NOMA allows users to share the same resources and distinguishes them in power, code, or some other emerging domains. In this subsection, we will introduce methods for several prominent NOMA schemes, including PD-NOMA, code-domain NOMA (CD-NOMA), and pattern division multiple access (PDMA). Then, we focus on methods dealing with challenges including asynchronous NOMA and NOMA with imperfect CSI.

\subsubsection{Practical NOMA Schemes}
\begin{itemize}
  \item \textit{PD-NOMA} is realized in the power-domain \cite{liu2017non}, in contrast to the traditional multiple access schemes relying on the time, frequency, code-domain or on their combinations. Users send messages using different power depending on their respective channel quality after classic channel coding and modulation. Multiple users share the same time-frequency resources, and then multiuser detection algorithms, such as SIC, are used at the receiver to detect different user signals from the superimposed signals. In this way, the SE can be enhanced at the cost of increased receiver complexity compared to conventional OMA. With the rise of multi-antenna technology, there is an increasing number of studies that combine PD-NOMA with multi-antenna to further improve SE \cite{7542118,9708418,9779941,choi2016power,liu2022effective}. In the multi-antenna scenario, the user clustering, beamforming design, and power allocation subject to the quality of service constraints need to be simultaneously considered and jointly optimized. For instance, the authors of \cite{liu2022effective} proposed a two-step user clustering and power control algorithm. Considering the intra-cluster interference and inter-cluster interference, K-means algorithm was used for user clustering, and then power control was done. Simulation results showed that user clustering by K-means can achieve lower energy consumption and higher EE.
  \item \textit{CD-NOMA} is inspired by CDMA systems, in which users share the same time-frequency resources but adopt unique user-specific spreading sequences. User-specific scrambling/interleaving, symbol spreading and symbol to resource mapping could be used to distinguish different users at the transmitter side \cite{chen2018toward}. Low-density signature (LDS)-CDMA \cite{hoshyar2008novel} uses a specially designed non-orthogonal spreading sequence with sparse characteristics. Each user extends its own information according to the signature sequence to the corresponding code chip for transmission. At the receiver, a near-optimum chip-level iterative multiuser decoding based on the message passing algorithm (MPA) is proposed to approximate optimum detection by efficiently exploiting the LDS structure. LDS-OFDM \cite{hoshyar2010lds} retains the benefits of OFDM-based multicarrier transmissions such as intersymbol interference avoidance and operates at a lower complexity than the optimal maximum a posteriori (MAP) detector of LDS-CDMA. Sparse code multiple access (SCMA) \cite{nikopour2013sparse} directly maps the bits to multidimensional codewords in the complex domain after channel coding at the transmitter, and the transmitted codewords of different users are non-orthogonally superimposed in a sparse spreading manner on the same time-frequency resources, and this sparse characteristic is used at the receiver for low-complexity multiuser joint detection. It is crucial to design the spreading sequences or the codewords in CD-NOMA. A set of user-specific non-orthogonal binary spreading sequences for uplink grant-free NOMA is studied in \cite{yu2020binary} to improve the performance of the joint channel estimation (CE) and multi-user detection. An improved low-density spreading sequence design based on projective geometry is proposed in \cite{millar2022low}. In \cite{chen2021reinforcement}, reinforcement learning was exploited to design the SCMA codebooks, which generates codebooks with superior quality and has a low complexity. Another breakthrough to achieve ow-complexity CD-NOMA  is rearranging OFDM subcarrier  sequences in a symbol  to efficiently supports multiuser access of lightweight IoT nodes \cite{295613}.
  \item \textit{PDMA} \cite{chen2016pattern} relies on uniquely designed multi-user diversity patterns to recognize non-orthogonal transmissions in the power, time, frequency, spatial and code domains. PDMA considers the joint design of the transmitter and receiver. On the transmitter side, transmitted data is mapped to a resource group that can consist of time, frequency, and spatial resources, or any combination of these resources, which is called PDMA pattern and can be represented by a binary vector. The pattern is important for separating the signals of users sharing the same resources, and its design should take into account the complexity of the detection. The general SIC algorithm is used at the receiver side to perform multiuser detection, to differentiate overlapping user signals based on different user patterns. 
\end{itemize}

\subsubsection{Asynchronous NOMA} 
As discussed in the previous section, the capacity region of the asynchronous uplink channel is larger than that of the synchronous uplink channel. As such, the burden of perfectly synchronizing different users does not apply to NOMA. In addition, it is beneficial to add intentional time asynchrony to improve the performance of NOMA systems.  

\textbf{Uplink ANOMA:}
Uplink ANOMA systems can be created by adding intentional timing mismatch among the symbols of different users. Assuming the timing mismatch is known, over-sampling generates sufficient statistics and provide sampling diversity. 
ANOMA offers a larger capacity region, compared to synchronous NOMA,  but requires dealing with IUI and ISI. A simple SIC is not optimal anymore and maximum-likelihood sequence detection is needed to manage the ISI \cite{MGHJ16}. One approach to manage ISI in ANOMA is to represent the ANOMA system with a Toeplitz system model and apply the trellis-based detection method to the corresponding Ungerboeck model for ISI channels. Then, as proposed in \cite{MGXZHJ21}, the channel diagonalization and turbo principle can be used to design an ANOMA transceiver. The optimal timing mismatch for a system with two users is half of the symbol interval. 
For systems with $K$ users, a good timing mismatch choice among different users is the symbol interval divided by $K$. 
The operational rate point for two users can be outside of the synchronous uplink channel's capacity region, i.e., it will outperform any possible synchronous NOMA system. In \cite{XZBHHJ19}, the sum throughput of two-user ANOMA systems is calculated  analytically. It is shown that by intentionally adding a timing mismatch, equal to half of the symbol interval between the symbols of the two users and over-sampling, the performance of ANOMA is always better than that of synchronous NOMA for a sufficiently large frame. Even when the exact timing mismatch between the two users is not known at the receiver, using an arbitrary equal-distance over-sampling, ANOMA's sum throughput is still more than that of the  synchronous NOMA. 
The average symbol error rate of ANOMA using QAM over Rayleigh fading channels is analytically calculated in \cite{9784923}. Also, an iterative detection method that takes into account the correlated nature of the noise and the error propagation is proposed. The approach is more effective for lower-order QAMs. 
In \cite{8669846}, another iterative method using message passing detection for ANOMA is designed and its performance is analysed. 
A massive device-to-device network utilizing ANOMA to  reduce the decoding complexity, energy consumption, and bit error rate is considered in \cite{10018312}. To manage co-channel interference, an optimization framework is defined to maximize the number of successfully decoded symbols.

\textbf{Downlink ANOMA:}
The main idea behind downlink ANOMA is to add intentional sub-symbol time delays between the symbols of different users before implementing the superposition coding at the transmitter. At the receivers, over-sampling provides sufficient
statistics for decoding. Because of the over-sampling, the noise is correlated and needs to be whitened. Singular-value decomposition and precoding can be used to create independent parallel channels that do not need SIC \cite{cui2017asynchronous}. In fact, adding intentional time delays will induce frequency-selectivity that can improve the performance in conjunction with beamforming and precoding methods \cite{8986830}. The performance of ANOMA, with or without precoding, is better than that of synchronous NOMA.  

\par
In NOMA, the stronger user decodes the signal of the weaker user and cancel its interference. Since the stronger user has access to the data from the weaker use, it can forward it to the weaker user, acting as a relay. Then, the weaker user has access to two copies of its symbols and can use them to improve the performance. In \cite{7222419}, the achievable average rate of such a cooperative NOMA (C-NOMA) scheme is analyzed. In \cite{7219393}, C-NOMA is designed for multiple antennas. Using MRC at the receivers and antenna selection at the transmitter, the outage probability is calculated analytically. It is shown in \cite{8910570} that, similar to a regular NOMA, adding intentional time delays to C-NOMA schemes will improve the performance. The throughput of both users in such a C-ANOMA system is calculated and it is shown analytically that C-ANOMA outperforms C-NOMA. In addition, the optimal timing mismatch is half of the symbol length for large frame sizes. In \cite{10006363}, a system similar to C-ANOMA is used for satellite communication. An optimization problem to jointly allocate resources to improve the fairness is defined and solved for a multi-satellite system.

\subsubsection{NOMA With Imperfect CSI}
While NOMA can provide many advantageous over OMA, the benefits of NOMA are usually shown under perfect assumptions, such as the availability of perfect CSI at both transmitter and receiver. For the multi-user/multi-cell environments for which NOMA is an essential technology, acquiring perfect CSI is not an easy task (if not impossible). Thus, designing systems to deal with imperfect CSI is very important when shifting from the OMA paradigm to a NOMA paradigm. 
Channel state information plays a crucial role in wireless communications. In a single-user system, communication system design heavily depends on how much channel information is available \cite{HJ05}.
Although the effect of imperfect CSI at the receiver on emerging NOMA techniques is still an ongoing research topic, similar to OMA systems, the estimated CSI can be modeled as the perfect CSI plus a Gaussian residual error. The achievable rates in NOMA depend on not only the individual estimates of channel coefficients, but also their relation. In uplink NOMA, the BS first decodes the signal of the strong user and then, after removing its interference, decodes the signal of the weaker user. In downlink NOMA, the stronger user decodes the weaker user’s signal first and after removing it, decodes its own signal. On the other hand, the weaker user only decodes its own message considering the other user’s signal as noise. Therefore, the decoding order which depends on the relative relation between channel coefficients plays a significant role in system performance. If the imperfect CSI at the receiver results in incorrect decoding orders, it can have catastrophic consequences. 

\par
Assuming perfect CSI at the receiver, the estimated channel values need to be communicated with other nodes. In some scenarios, channel reciprocity is used for this purpose. However, in most practical cases, the estimated channel values should be quantized and sent back to other nodes through a limited feedback channel. 
Downlink NOMA with limited feedback is analyzed in \cite{7968348}, where variable-length quantizers that can approach the perfect CSI performance are designed. It is proved that the relationship between the feedback rates and the losses in rate and outage probability is at least exponential. 
In \cite{8938127}, downlink NOMA systems in which the transmitter does not know the perfect CSI is studied. It is shown that  if the power allocation strategy is not well-designed, outages can occur. In addition, for specific target rates, a power allocation method that achieves the OMA outage performance is designed. 
In \cite{8999638}, the performance of a limited feedback system using NOMA in a hybrid UHF/mmWave downlink network is analyzed. The achieved outage probability is much lower than that of a network with no feedback.
In \cite{9103094}, optimal quantizer algorithms for NOMA and ANOMA are designed. A two-user downlink ANOMA
system with limited feedback is considered and its max-min rate's upper and lower bounds are expressed in closed-form. For the same max-min rate, the required feedback rate of ANOMA is less than that of NOMA. 
The design of a limited feedback that includes reconfigurable intelligent surfaces in NOMA systems is considered in \cite{MAHJ23}. The rate loss resulted from the quantization error is calculated. 

\subsection{Machine Learning for NOMA}
The capability of DL can greatly aid in resolving diverse complex problems in NOMA such as CSI acquisition \cite{8387468}, signal constellation design \cite{8766718}, resource allocation, user pairing and multiuser detection. Generally, two categories of learning techniques have been applied in the literature to solve the above problems, including DL and reinforcement learning (RL). DL has been widely used for estimating CSI \cite{gui2018deep}, throughput prediction \cite{vu2021performance}, transceiver design \cite{8625480,8952876}, and resource management \cite{9123600,9314919}. Based on the agent-learning control mechanism, RL and deep RL (DRL) are used in resource management, especially in dynamic resource management \cite{8927898,9381701,9686696}. 

\par
Resource allocation is essential in NOMA in order to achieve the optimal total sum rate. Traditional methods try to maximize the EE or the sum rate under a series of constraints. Most optimization problems are nonconvex and cannot be equivalently transformed into convex optimization problems, so it can only be solved suboptimally using methods such as successive convex approximation (SCA) \cite{nedic2018parallel}, which has been used in power allocation. In \cite{9123600}, a DL-based framework for subchannel and power allocation was proposed to improve NOMA EE in mmWave networks. The subchannel allocation was solved by semi-SL, in which some labeled data was initially generated by a two-sided matching scheme. Using the generated data along with unlabeled data, the network was trained and was able to accurately predict the real label of an unlabeled input. The power allocation was solved using a DL scheme, which replaces the complex iterative process and only uses simple multiplication and addition of matrices to compute the output. The simulation results showed that the proposed framework achieved a higher EE and a lower complexity. Considering a relay-aided device to device system, a deep neural network (DNN)-based algorithm was designed in \cite{9314919} to replace intractable coupled joint optimization and acquire a joint power loading solution at the source and relay nodes. Their results revealed that the DNN model leads to promising performance in terms of sum rate and computational complexity. In \cite{10064048}, a two-step algorithm was designed to perform sub-band assignment and power allocation successively for each NOMA-based IoT user to minimize the transmit power sum. And unsupervised learning (USL) was implemented for power control. 

\par
In addition to DL methods, RL was applied for dynamic resource allocation in time-varying communication channels \cite{9686696,ling2023dqn,zheng2021channel,9964376,8927898,9381701}. In \cite{9964376}, a model-free centralized and distributed approach based on DRL was proposed in multiband hybrid OMA-NOMA to jointly optimize user association, transmit power allocation, sub-channel assignment, and multiple access technique selection. In \cite{8927898}, a two-step DRL-based algorithm was proposed to solve the dynamic optimization problem of EE. To be specific, using the current channel conditions as input, a deep q-network (DQN) was designed to obtain the optimal subcarrier allocation policy, and a deep deterministic policy gradient (DDPG) network was used to dynamically output the transmit power for all users. To maximize the average performance of sum rates in uplink NOMA-IoT, two RL algorithms, i.e., SARSA-learning and DRL, were proposed in \cite{9381701} to dynamically allocate users and balance resource blocks and network traffic. Numerical results demonstrated that the two algorithms have a lower complexity and outperformed OMA systems. In \cite{9686696}, DQN and DDPG were proposed for sub-carrier assignment and power allocation in RIS assisted semi-grant-free NOMA transmission, respectively. Two DDPGs were integrated to assign amplitude and phase shift to RIS's reflecting elements. A distributed cooperative channel assignment
and power control approach based on multi-agent RL is presented in \cite{9311792} to solve the massive access management problem.

\par
The design of the transceiver is another important issue in NOMA. In practice, there are multiple practical schemes, e.g., PD-NOMA and CD-NOMA, and the signal formats of each scheme are significantly different, requiring different transceiver designs, which hinders the standardization and implementation of NOMA. The DL method emerges and makes possible a unified transceiver architecture. In \cite{8254356}, a data-driven model was proposed for SCMA. The traditional decoding algorithm in SCMA, such as MPA, was replaced by a fully connected DNN (FC-DNN), which is trained using a synthetic dataset to minimize the symbol error rate. It also reduced the computation time to a large extent. The variational autoencoder (VAE) was exploited to perform signal detection with jointly designed spreading signatures \cite{8625480}. An end-to-end learning-based SCMA framework was introduced in \cite{luo2022novel} to jointly design the codebooks and low-complexity decoder. The above data-driven DNNs lead to a unified framework for different NOMA schemes, while being trained using different datasets. However, data-driven methods do not take advantage of the structure of NOMA signals and thus lead to low training efficiency. Model-driven methods bring structured DNN based on traditional algorithms and domain knowledge. In \cite{8827912,9130955}, an SIC-inspired DNN was designed for multi-user detection. The network was inspired by the conventional SIC detection structure, in which DNNs replaced the detection layer in SIC, and the output of the first branch was fed to the second branch to cancel inter-user interference like SIC. Numerical results demonstrated the effectiveness and superior performance of the model-driven method. The DL methods mentioned above consider block-wise optimization. A unified end-to-end optimization framework called DeepNOMA was proposed in \cite{8952876}, which consists of a channel model, a multiple access signature mapping module, and a multiuser detection model. End-to-end optimization is enabled by minimizing the overall reconstruction loss. What is more, each users' signals in non-orthogonal transmissions are treated as multiple distinctive but correlated tasks, and multitask learning and balancing techniques are used to improve the performance, guarantee fairness among users, and avoid local optima.

\par
In NOMA, there are also other important problems that can be solved efficiently by DL. User pairing is essential in NOMA systems, as a reasonable user pairing strategy can improve system performance. In \cite{9417564}, after obtaining the power allocation strategy, a user pairing matrix was designed based on the channel gain of all users and a DQN-based algorithm was utilized to solve the user pairing problem to get the maximum sum rate. Reference \cite{9145383} expanded the scenario to multicell NOMA and jointly optimized the user pairing and association problem. The pointer network (PtrNet) was used to solve the combinatorial optimization problem, and DRL was utilized at the training phase. Simulation results showed that the scheme achieved near-optimal performance, which is better than the random user pairing and association heuristic by up to 30\%. A DL-based adaptive user pairing scheme, which decides the two optimal user pairings, was proposed in RIS-aided MIMO-NOMA to maximize the overall spectral efficiency in \cite{10129167}.

\par
Machine learning has also been leveraged to improve transmission security in NOMA networks. In \cite{li2020cache}, the interaction between the source station and the unmanned aerial vehicle (UAV) was viewed as a dynamic game and a DRL algorithm was applied for dynamic power allocation at the source station to suppress the attack motivation of the UAV. The simulation results showed that this strategy improves the system data rate and suppresses the attack probability. The NOMA transmission game against jamming was investigated in \cite{xiao2017reinforcement}, where a fast Q-based NOMA power allocation scheme that combines the hotbooting technique and Dyna architecture is proposed for a dynamic game to accelerate the learning and thus improve the communication efficiency against smart jamming. The summary of the existing work on machine learning for NOMA is given in Table \ref{Summary-ML-NOMA}.

\begin{table*}[htb]  
\renewcommand\arraystretch{1.2}
\begin{center}   
\caption{Summary of Existing Work on Machine Learning for NOMA}  
\label{Summary-ML-NOMA} 
\begin{tabular}
{|m{2.3cm}<{\centering}|m{2cm}<{\centering}|m{1.8cm}<{\centering}|m{8.3cm}|m{1.6cm}<{\centering}|}
\hline
\textbf{Category}  & \textbf{Scenarios}  & \textbf{Algorithm}    & \textbf{Characteristics} & \textbf{Ref.}       \\ \hline
\multicolumn{1}{|c|}{\multirow{8}{*}{Resource allocation}} & mmWave NOMA   & DNN   & Use semi-SL for subchannel allocation                 & \cite{9123600} \\ \cline{2-5} 
\multicolumn{1}{|c|}{}   & Relay-aided NOMA   & DNN   & Propose a joint power loading solution at source and realying nodes   & \cite{9314919}  \\ \cline{2-5} 
\multicolumn{1}{|c|}{}   & NOMA-IoT   & USL-based DNN   & Implement USL for power control   & \cite{10064048}  \\ \cline{2-5} 
\multicolumn{1}{|c|}{}   & NOMA-aided MEC & DQN  & Use DQN-based offloading strategy to minimize the system cost  & \cite{ling2023dqn}   \\ \cline{2-5} 
\multicolumn{1}{|c|}{}   & Hybrid NOMA   & Actor-critic  & Use proximal policy optimization and recurrent neural networK & \cite{zheng2021channel}    \\ \cline{2-5} 
\multicolumn{1}{|c|}{}  & Multi-band hybrid OMA-NOMA  &  DRL & Propose model-free centralized and distributed approaches based on DRL & \cite{9964376}    \\ \cline{2-5} 
\multicolumn{1}{|c|}{}   & RIS assisted NOMA  & DQN, DDPG   & Propose a two-step DRL based aligorithm  & \cite{8927898,9686696}  \\ \cline{2-5} 
\multicolumn{1}{|c|}{}  & NOMA-IoT  & SARSA-learning,DRL & Design novel 3D state and action spaces    & \cite{9381701}     \\
\cline{2-5} 
\multicolumn{1}{|c|}{}  & NOMA-IoT  & DQN & Build a QoS-aware reward to cover both network EE and devices QoS   & \cite{9311792}     \\
\hline
\multirow{3}{*}{Transceiver design} & \multirow{3}{*}{Grant-free NOMA} & FC-DNN     & Propose a unified NOMA framework    & \cite{8254356,8625480,luo2022novel} \\ \cline{3-5}     &     & SIC-inspired DNN  & Design a two branch network and the structure is SIC-liked     & \cite{8827912,9130955} \\ \cline{3-5} 
   &   & DeepNOMA   & Propose an end-to-end optimization framework  & \cite{8952876} \\ \hline
\multirow{3}{*}{User pairing}    & NOMA   & DQN  & Design and optimize a user pairing matirx    &\cite{9417564}   \\ \cline{2-5} 
   & Muticell NOMA    & PtrNet & Use a reinforce-based method to perform parameter optimization   & \cite{9145383}  \\ \cline{2-5}
   & RIS-aided MIMO-NOMA   & DNN  & Propose DL-based adaptive user pairing scheme    &\cite{10129167}\\ \hline
\multirow{2}{*}{Secure issue}   & NOMA  & Q-learning  & View the interaction between BS an UAV as dynamic game    & \cite{li2020cache}    \\ \cline{2-5} 
  & MIMO NOMA     & Hotbooting Q-learning & Apply Dyna architecture and hotbooting techniques    & \cite{xiao2017reinforcement}  \\ \hline
\end{tabular}
\end{center}   
\end{table*}
\renewcommand\arraystretch{1}

\section{Interplay With Other Next-Generation Wireless Technologies and Scenarios}
Having introduced the fundamentals and advanced signal processing and learning techniques of MRA and NOMA, in this section, we focus our attention on the promising interplay with other new next-generation technologies and scenarios, e.g., AoI, NFC, ISAC, STARS and semantic communications. 

\subsection{NGMA with New Scenarios}
In the future, with the rapid development of IoT, the diversity of IoT devices will increase, leading to varying communication needs among different devices. For example, in the context of Internet of Vehicles, in addition to the massive access characteristic of MMTC, it is also necessary to ensure low latency and high reliability for certain sensors. In this subsection, we introduce the consideration of AoI and heterogeneous services.

\subsubsection{Age of information}
Upcoming use cases and applications will soon impose new and stricter demands on MRA. In the design of 6G systems, new KPIs will become pivotal, e.g., AoI. In future 6G networks, information freshness will be increasingly important for applications of various types of real-time monitoring systems (e.g., smart driving) and state update systems (e.g., industrial control). The use of outdated information could profoundly compromise the precision and dependability of systematic decision-making processes. To quantify information freshness, a new metric, i.e., AoI, is introduced to measure the information freshness at the receiver \cite{5984917}. To reduce the AoI of the system, an optimized age based random access scheme is proposed in \cite{mlsp2024}, and a neural network where AoI is aided as the prior information is further proposed. To improve the overall status update performance, a joint access control and resource allocation scheme is proposed in \cite{9181539}, taking AoI as the performance metric. To achieve a low average AoI and large throughput in satellite-based IoT, a grant-free age-optimal random access protocol is proposed in \cite{9766119}, where the number of access time slots in each transmission frame is optimized. In \cite{10039352}, an RL-based transmission strategy is proposed in which the AoI of the users and the total throughput of the system are jointly optimized. Based on the AoI levels of users, they are divided into two groups and assigned distinct access patterns. According to \cite{9377549}, in the slotted-ALOHA framework, a device is allowed to access the channel with a constant probability if its AoI exceeds a specified limit. The long-term expected AoI is calculated, followed by an optimization of the access probability and the threshold.

\subsubsection{Heterogeneous services}
In future wireless networks, numerous distinct application scenarios are encountered. In these contexts, users with different Quality of Service (QoS) needs must coexist. Emerging in heterogeneous radio access networks, a blend of enhanced mobile broadband (eMBB), MMTC, and ultra-reliable low-latency communication (URLLC) devices could be accommodated within the same spectrum resource. In \cite{9779514,10102680,9843911}, the coexistence of eMBB and URLLC is studied. The joint resource allocation problem to maximize the minimum expected achieved rate of eMBB users while satisfying the low latency and high reliability constraints of URLLC users is studied in \cite{9779514}. Furthermore, URLLC traffic is overlapped on eMBB by adopting the NOMA superposition technique. In \cite{10102680}, the non-orthogonal coexistence between eMBB and URLLC in the downlink of a multi-cell massive MIMO system is rigorously analyzed. Puncturing and superposition coding are considered as alternative downlink coexistence strategies, and the eMBB spectral efficiency and the URLLC error probability are analyzed. In \cite{9843911}, the optimal frequency and power allocation scheme and eMBB-URLLC pairing policy is derived for scheduling URLLC traffic in a downlink system in the presence of eMBB traffic. A preamble allocation scheme using hierarchical reinforcement learning is proposed for the coexistence scenario of MMTC and URLLC in \cite{10238402}. A promising scheme based on RSMA for network slicing is proposed in \cite{10190330} to guarantee the performance of heterogeneous devices. Analysis shows that RSMA can outperform its NOMA counterpart, and obtain significant gains over OMA in some regions. In \cite{10171175}, a novel hierarchical hybrid access class barring (ACB) and backoff (BO) scheme is proposed, where the hybrid ACB-BO is exploited to balance the delay-energy tradeoff, and the hierarchical structure is proposed to prioritize communication services. Deep reinforcement learning is applied to the proposed random access scheme to dynamically adjust the ACB factors and BO indicators in an online manner.

\subsection{Near-field Non-Orthogonal Multiple Access}
In 6G, paradigm shifts are currently taking place in the architecture to satisfy the stringent requirements. Specifically, by increasing the aperture size of transceivers and using extremely high-frequency bands, the achievable data rate and connectivity can be massively improved. In the meantime, near-field signal propagation in future wireless networks will become important, which brings new opportunities for NOMA. In the following, we first introduce the key features of NFC. Then, we highlight the \textit{beamfocusing} property in NFC which can benefit NOMA. Finally, we introduce the beamfocusing-empowered NOMA communications.

\subsubsection{Key Features of NFC} 

In wireless communication, the \textit{near-field} region refers to the area close to the transceivers so that the electromagnetic field in that region exhibit near-field effects. One widely used rule-of-thumb for determine the near-field region is the \textit{Rayleigh distance}, which depends on the aperture size of the transceivers and the carrier frequency. The larger the aperture size and the higher the carrier frequency, the larger the near-field region becomes.
Within the near-field region, signal propagation has different characteristics compared to that within the far-field region. In the following, we elaborate on the new features of NFC.
\begin{itemize}
    \item \textbf{Distance-Dependant Radiation Pattern}: 
The near-field region in wireless communication deviates from the free-space path loss observed in far-field communications. In particular, the near-field radiation pattern, which represents the angular distribution of signal power, exhibits variations as a function of the communication distance. This means that even in a single direction, the path loss may oscillate instead of following a monotonically decreasing trend, as described by the far-field Friis' path loss formula \cite{10220205}. As we will elaborate in Section.~\ref{NF-beamfocusing}, this near-field property gives rise to the possibility of \textit{beamfocusing}.
    \item \textbf{Spherical and Irregular Wavefronts}: In contrast to the planar waves in the far-field region, near-field radiation has a spherical or irregular wavefront. Close to the Rayleigh distance, the wavefront appears spherical, with the energy spreading equally in all directions. For regions close to the transmitter, the wavefront can become irregular or distorted. This is because the antenna's physical structure and geometry can introduce variations in the propagation path and cause interference effects.
    \item \textbf{Faster Decay and Reactive Field}:
    Near-field also exhibits a faster decay due to the dominance of spherical waves and evanescent waves. Evanescent is a type of electromagnetic field whose energy decrease exponentially with distance. The region where evanescent wave dominates is referred to as the reactive near-field. It is important to note that the reactive near-field is confined to just a few wavelengths from the antennas or antenna arrays, even if they are infinitely large \cite{10614327}. As a result, this region is typically disregarded in near-field communication and signal processing design.
    \item \textbf{Enhanced Degrees-of-Freedom (DoFs)}: The line-of-sight (LoS) MIMO channel in NFC typically exhibits a higher number of DoFs. In contrast to the LoS MIMO channel in conventional far-field communications, where the achievable DoF is 1, the enhanced spatial multiplexing gain can be achieved in NFC even without a rich scattering environment, which is the major benefit of NFC.
    \item \textbf{Non-Uniform Pathloss}: The communication channel in NFC can exhibit high variability due to quick changes in the relative positions and orientations between the transmitter and receiver. As a result, the subchannels between two multi-antenna transceivers exhibit non-uniform pathloss. In channel modeling, non-uniform spherical wave models should be used characterize these near-field channels.
\end{itemize}

\subsubsection{Near-field Beamfocusing Property}\label{NF-beamfocusing}

\begin{figure}[t!]
    \centering
    \captionsetup{font={footnotesize}}
    \begin{subfigure}[t]{0.24\textwidth}
        \centering
        \includegraphics[width=1\textwidth]{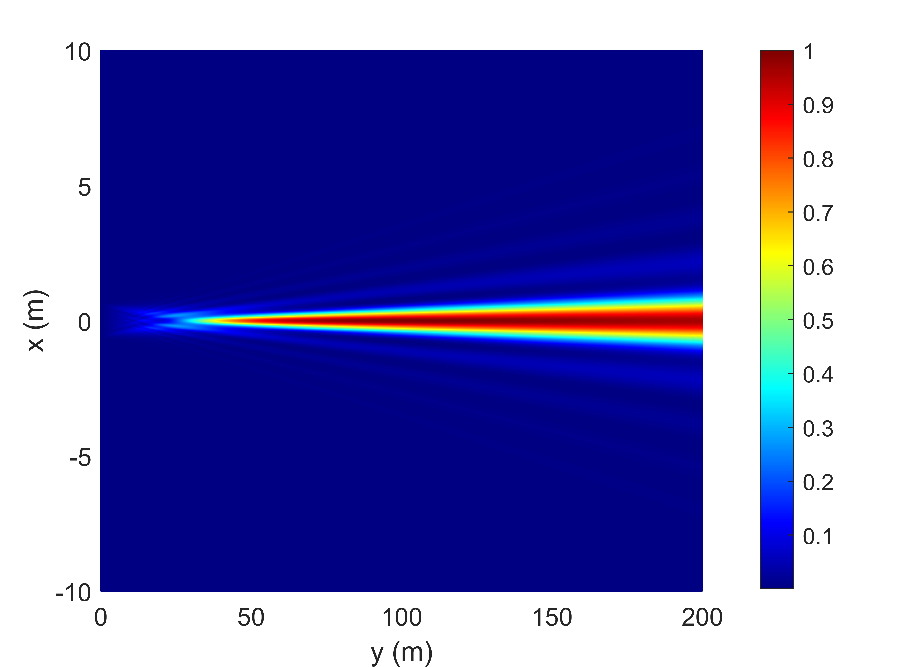}
        \caption{Far-field beamsteering.}
        \label{fig:ff_steering}
    \end{subfigure}
    \begin{subfigure}[t]{0.24\textwidth}
        \centering
        \includegraphics[width=1\textwidth]{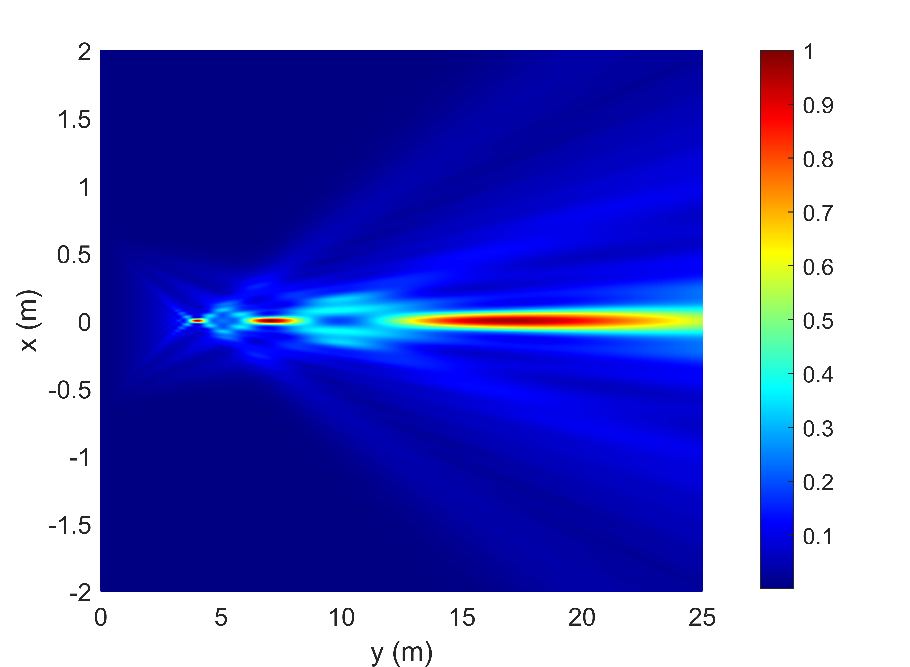}
        \caption{Near-field beamfocusing.}
        \label{fig:nf_focusing}
    \end{subfigure}
    \caption{Beampattern of far-field beamseering and near-field beamfocusing.}
\end{figure}

As previously discussed, the signal propagation in the near-field region exhibits distinct characteristics, primarily manifesting as spherical waves, in contrast to the planar waves observed in conventional far-field communications. Consequently, the fundamental principles underlying beamforming techniques diverge between near-field and far-field communication scenarios within multi-antenna systems. Specifically, as shown in Fig. \ref{fig:ff_steering}, in the far-field region, the angular field distribution remains unaffected by the distance from the transmit antenna array, resulting in the capability of the far-field beamforming process to solely steer towards a predetermined direction, commonly denoted as \emph{beamsteering}. Conversely, in the near-field region, the angular field distribution is contingent upon the distance from the transmit antenna array. Thus, as shown in Fig. \ref{fig:nf_focusing}, the near-field beamforming technique exhibits the ability to concentrate the beam energy at a specific distance along a designated direction, referred to as \emph{beamfocusing}. 

Note that near-field beamfocusing is not a new research topic and can date back more than 60 years \cite{bickmore1957focusing, sherman1962properties, musil1967properties}. However, it was mainly investigated in the field of physics and antenna propagation. In recent years, there has been a notable trend towards the utilization of extremely large-scale antenna arrays (ELAAs) and ultra-high frequencies, such as mmWave and Terahertz (THz) bands, to meet the rigours communication requirements of future wireless systems. Consequently, the near-field region in wireless systems becomes non-negligible, with its spatial extent reaching tens or even hundreds of meters with respect to the BS \cite{10220205}. Given this very large near-field region, many researchers began to investigate the potential benefits of near-field beamfocusing in wireless communications \cite{bjornson2021primer, zhang2022beam, 10123941}. Specifically, the basic principle of near-field beamfocusing in wireless communications was discussed in \cite{bjornson2021primer}, where several distances distinguishing the near-field and far-field regions of an antenna array were defined. Considering a multi-user communication scenario, the authors of \cite{zhang2022beam} investigated the beamfocusing capability of different antenna architectures. It was demonstrated that in the near-field region, different beams can be generated to focus on different communication users located in the same direction without causing significant inter-user interference. As a further advance, a novel concept of location-division multiple access (LDMA) was proposed in \cite{10123941}. By exploiting the additional distance dimensional provided by near-field beamfocusing, multiple users can be supported in the location domain in the near-field region, which provides much higher DoFs than the angular domain in the far-field region.
\begin{figure}[t!]
    \begin{center}
        \includegraphics[scale=0.6]{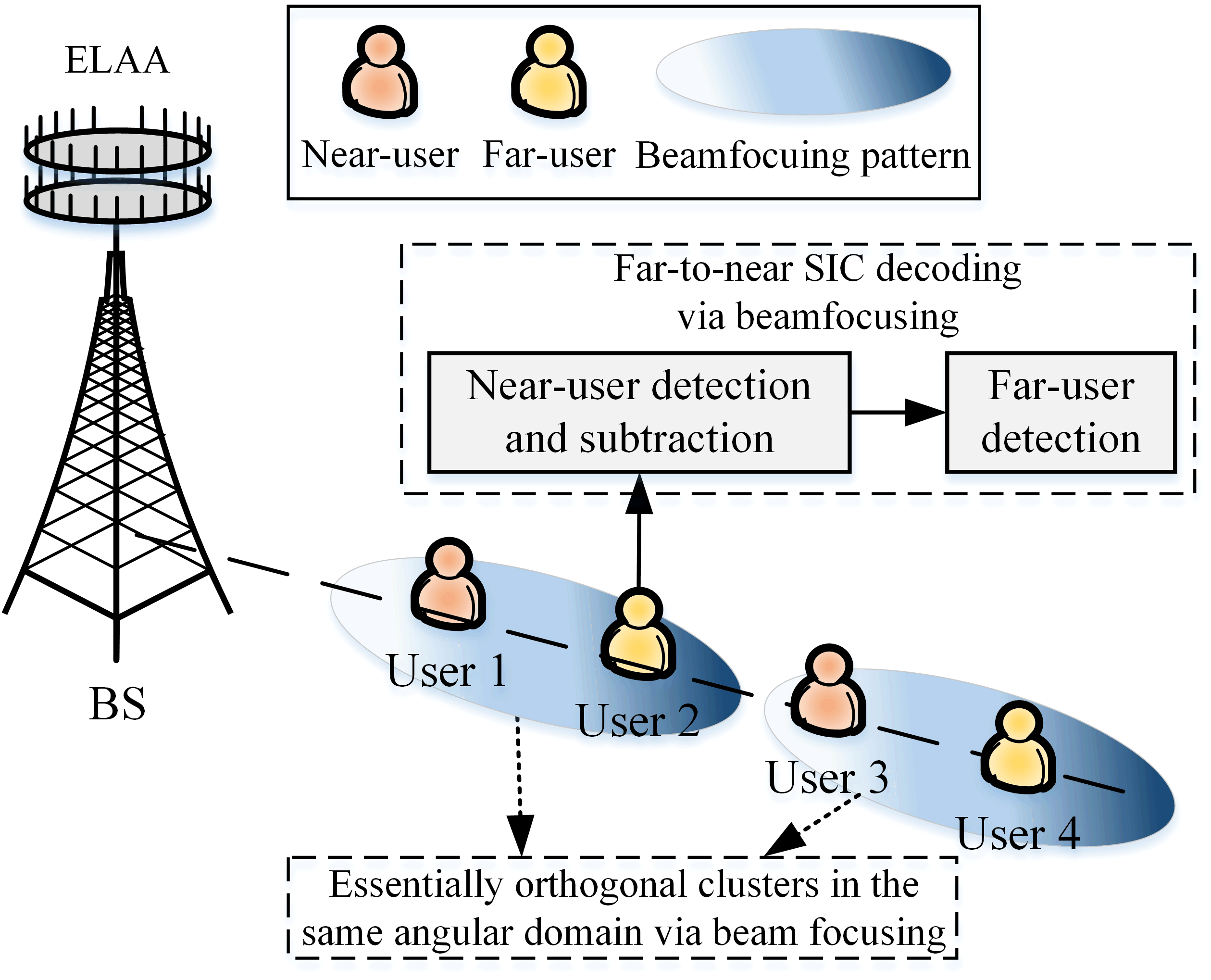}
        \caption{Illustration of near-field beamfocusing-empowered NOMA, where the near-field beamfocusing vectors divide users in the same angular direction into two clusters and the signal energy is focused on the far-user in each cluster.}
        \label{NFC_NOMA}
    \end{center}
\end{figure}
\subsubsection{Near-field Beamfocusing-empowered NOMA} With the aid of the promising beamfocusing property in NFC, there are two main benefits that can be achieved by the near-field beamfocusing-empowered NOMA.
\begin{itemize}
    \item \textbf{Far-to-near SIC decoding}: With the properly designed beamfocusing vectors at the BS, the signal can be more focused on the users far from the BS than those near-users, i.e., the effective channel gain of the far-user can be larger than that of the near-user. As a result, it enables the far-user to carry out SIC to detect and remove the signal of the near-user, thus realizing the far-to-near SIC decoding order, as shown in Fig. \ref{NFC_NOMA}. This is almost impossible to realize for conventional far-field NOMA communications, where the SIC decoding order among users is generally determined by their distance to the BS. Therefore, the near-field beamfocusing-empowered NOMA provides enhanced flexibility in the SIC decoding order design, which helps to better satisfy the communication requirements of each user.
    \item \textbf{Distance-domain user clustering}: Moreover, with the aid of beamfocusing, users located in the same angular domain can be further distinguished via the distance domain into multiple (nearly) orthogonal clusters, as shown in Fig. \ref{NFC_NOMA}. This is greatly different from the far-field NOMA, where the user clustering is mainly determined by angular similarity. By doing so, on the one hand, the inter-user interference can be further mitigated due to the orthogonality between user clusters. On the other hand, with the reduced number of users in each cluster, the number of SIC operations reduces accordingly, which reduces the complexity of using NOMA.
\end{itemize}
To exploit the above benefits of the near-field beamfocusing-empowered NOMA, two novel frameworks were proposed in \cite{jiakuo} for a BS employing the hybrid beamforming structure, namely, single-location-beamfocusing near-field NOMA and multiple-location-beamfocusing near-field NOMA. For the single-location-beamfocusing near-field NOMA, users located in the same angualr direction but having distinct communication requirements are clustered into one group, which are served by one analog beamformer focusing on one specific user. For the multiple-location-beamfocusing near-field NOMA, users in the same group can have different angular directions, which is served by one analog beamformer focusing on multiple users by exploiting the beam-splitting technique. It shows that compared to conventional far-field NOMA, the proposed near-field beamfocusing-empowered NOMA can achieve higher spectral efficiency and enhanced inter-user interference mitigation. Moreover, the authors of \cite{10129111} investigated the NOMA-based coexistence of near-field and far-field communications, where the near-field beamfocusing vector is shown to be able to serve an additional far-field user via NOMA. Note that the investigation of near-field/hybrid-field NOMA communications is still in an early stage, more research efforts are required to unlock the full benefits.

\subsection{NOMA for Integrated Sensing and Communications}
ISAC has been recognized as a crucial facilitator of ubiquitous intelligence in 6G \cite{cui2021integrating, 10024901}. ISAC systems aim to integrate communication and sensing functionalities, leveraging shared spectrums, hardware platforms, and signal processing modules. Nonetheless, the coexistence of these two functions poses significant challenges in terms of interference cancellation and resource allocation within the ISAC framework. Consequently, the subsequent section presents a comprehensive examination of ISAC systems from a multiple-access perspective, while also exploring the potential advantages of NOMA in enhancing ISAC systems.

\subsubsection{ISAC from the Multiple Access Perspective}
As depicted in Fig. \ref{fig:sensing}, the ISAC system can be conceptualized as a multiple-access framework consisting of two distinct users: the communication user and the sensing user (target). The communication user operates in either an uplink (UL) mode, transmitting communication signals to the BS, or a downlink (DL) mode, receiving communication signals from the BS. In contrast, the sensing target behaves more like an inherent full-duplex user, concurrently “receiving” and “transmitting” signals. This distinction arises from the divergent operational mechanisms between sensing and communication. More specifically, while communication users are capable of actively engaging in coding and decoding processes to convey or recover information, the sensing target is passive. Consequently, the BS is responsible for transmitting probing signals to the sensing target and subsequently receiving the sensing echo signals reflected by the target. This enables the BS to acquire crucial parameters such as location, speed, and shape. It is worth noting that in some cases, such as the sensing-assisted beamforming framework described in \cite{liu2020radar}, the communication user and sensing target may be the same physical object, which is out of the scope of our discussion.

\subsubsection{Motivation of Employing NOMA for ISAC}
\begin{figure}[t!]
    \centering
    \captionsetup{font={footnotesize}}
        \begin{subfigure}[t]{0.24\textwidth}
        \centering
        \includegraphics[width=1\textwidth]{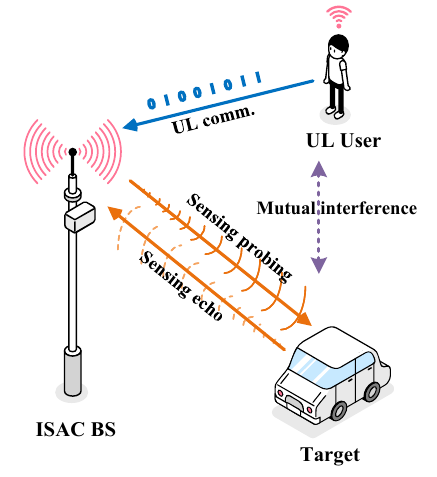}
        \caption{Uplink ISAC systems.}
        \label{fig:ul_sensing}
    \end{subfigure}
    \begin{subfigure}[t]{0.24\textwidth}
        \centering
        \includegraphics[width=1\textwidth]{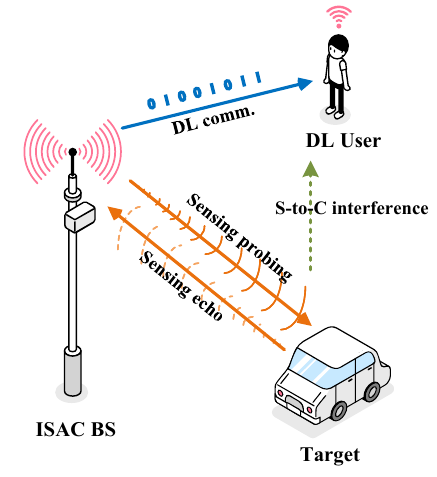}
        \caption{Downlink ISAC systems.}
        \label{fig:dl_sensing}
    \end{subfigure}
    \caption{Illustration of ISAC systems from the multiple access perspective.}
    \label{fig:sensing}
\end{figure}

Based on the above discussion, the concept of mitigating inter-user interference within multiple-access frameworks can be applied to solve inter-function interference in ISAC systems. A direct approach entails assigning “orthogonal” resource blocks in the frequency, time, or code domain to communication and sensing functions, resembling the concept of OMA where “orthogonal” resource blocks are allocated to different users. This orthogonal allocation strategy avoids inter-function interference in ISAC systems. However, while such an interference-free design reduces implementation complexity, it may result in inefficient resource utilization. As a remedy, non-orthogonal ISAC is a more popular design principle \cite{10024901}. In a manner analogous to NOMA, this approach allows for the joint utilization of resource blocks by both communication and sensing functions, resulting in a substantial improvement in resource efficiency. However, it is imperative to effectively mitigate the occurrence of inter-function interference. Therefore, in the following, we discuss the property of inter-function interference in non-orthogonal ISAC systems in both uplink and downlink non-orthogonal ISAC systems. 
\begin{itemize}
    \item \textbf{Uplink ISAC}: The uplink ISAC system is depicted in Fig. \ref{fig:ul_sensing}, where an uplink communication user and a sensing target are present. In this configuration, the sensing target can be considered as a virtual uplink communication user, “transmitting” echo signals to the BS. Consequently, the BS receives a superimposed uplink communication signal and sensing echo signal, leading to mutual interference between the communication and sensing functions.
    \item \textbf{Downlink ISAC}: The downlink ISAC system is illustrated in Fig. \ref{fig:dl_sensing}, involving a downlink communication user and a sensing target. Unlike uplink ISAC systems, the sensing target cannot be entirely regarded as a virtual downlink communication user in the downlink scenario. This is because it has been demonstrated that, in the downlink case, all communication signals can be utilized for sensing, while the sensing signals should be exclusively dedicated to achieving the full sensing DoFs \cite{liu2020beamforming}. Consequently, only sensing-to-communication interference is present in this case.
\end{itemize}
Given the resemblance between non-orthogonal ISAC and NOMA in terms of their underlying principles, it is natural to leverage NOMA techniques to address inter-function interference in ISAC. However, due to the inherent distinction between the original purpose of NOMA in enabling efficient communication and the requirements of ISAC systems, it becomes necessary to redesign NOMA specifically for ISAC, resulting in the development of \emph{multi-functional NOMA}. Consequently, numerous studies have emerged for studying NOMA-ISAC systems, which will be detailed in the following.

\subsubsection{Existing Works on NOMA-ISAC}
The performance of uplink NOMA-ISAC systems was initially explored in \cite{ouyang2022performance}, wherein the authors employed the SIC technique to mitigate inter-functional interference. Building upon this concept, the authors of \cite{wang2022noma_jsac} proposed a NOMA-aided protocol that facilitates the seamless integration of communication, sensing, and computing functions. It was demonstrated that the proposed NOMA-aided protocol outperforms the conventional protocol based on SDMA due to its superior interference cancellation capabilities. As a novel contribution, a semi-NOMA framework was conceived for uplink ISAC systems in \cite{10036107}. This framework involves the sharing of only a subset of resource blocks between the communication and sensing functions, thereby enabling a more adaptable resource allocation scheme and enhanced interference cancellation capabilities. Furthermore, a recent study \cite{memisoglu2023csi} proposed a model and pattern based environment sensing technique utilizing CSI, which can be executed concurrently with communication functions through NOMA techniques. 
Regarding downlink ISAC systems, the potential of NOMA was initially explored in \cite{9668964}. While focusing on NOMA within the communication function, this study demonstrated that NOMA can still offer additional DoFs for ISAC, particularly in scenarios with system overload or highly correlated communication channels. Building upon this foundation, the authors of \cite{wang2022inspired} extended the concept of NOMA to address the sensing-to-communication interference between the two functions. They proposed a NOMA-inspired ISAC framework that effectively eliminates sensing-to-communication interference and enables the simultaneous utilization of dedicated sensing signals for communication. Furthermore, the integration of the sensing function into multicast-unicast communication systems was effectively achieved with the aid of NOMA, as illustrated in \cite{mu2022noma}. To address security concerns in NOMA-ISAC systems, \cite{yang2022secure} introduced a secure precoding approach, offering a solution to the security problem. Lastly, the issue of sensing scheduling in NOMA-ISAC systems was investigated in the most recent work \cite{10129092}, which was jointly optimized with the beamforming. Some other works (e.g., \cite{chen2022beamforming}) also investigate the application of NOMA in full-duplex ISAC systems, where the UL and DL communication users coexist. It is important to note that, in addition to the resource allocation and interference management issues discussed in the aforementioned works, waveform design is also crucial in ISAC system development. Specifically, sensing functions typically require deterministic waveforms, while communication functions rely on random signals to convey information \cite{10147248}. Further research is needed to explore how NOMA techniques can be utilized to address this inherent tradeoff from the waveform perspective.

\subsection{Simultaneously Transmitting and Reflecting Surfaces (STARS) for NOMA}
In contrast to wireless networks ranging from the initial 1G to the latest 5G, which were primarily designed to overcome challenges posed by unpredictable radio conditions such as signal fading and blockages, RISs enable the establishment of a ``smart radio environment'' to facilitate the realization of 6G in a flexible and sustainable manner~\cite{ahead}. Significantly, the two-dimensional configuration and nearly passive operational characteristics of RISs contribute to their exceptional compatibility with prevailing wireless technologies. Recently, a novel type of STARS was proposed, which integrates both transmission and reflection (T\&R) functions into a single RIS and achieves full-space coverage of the smart radio environment. In the following, we will first give a brief introduction to STARS and then discuss the benefits of STARS-enhanced NOMA.  

\subsubsection{Basis of STARS}
At its core, STARS is a periodic structure with adjustable T\&R phase-shift coefficients. STARS builds upon the capabilities of RISs by introducing the capability to simultaneously transmit signals while reflecting them. This means that in addition to modifying the wireless channel by reflecting the incident signals, STARS can also let signals penetrate through the surface. By controlling the T\&R coefficients of the reconfigurable elements, STARS can dynamically adjust the phase and amplitude of both the transmitted and reflected signals.
The simultaneous transmissions and reflection capabilities of STARS offer several advantages in wireless communication systems. Firstly, they provide enhanced signal coverage and connectivity by simultaneously serving users on both sides of the surface. This enables improved communication links and increased network capacity.
Secondly, STARS can enable cooperative communication and multi-user interference management. By intelligently controlling T\&R coefficients, STARS can coordinate the signals of multiple users, manage interference, and optimize the overall system performance.
Furthermore, STARS is compatible with existing wireless technologies, making it a promising candidate for integration into future wireless networks. It can be seamlessly integrated with other communication technologies, including NOMA and MIMO, to further enhance network performance.
\\
In terms of hardware modeling, each element of a STARS can be treated as a lumped circuit with electric and magnetic impedances. The basic hardware and channel models for STARS were proposed in \cite{xu_star}. By configuring the values of two complex-valued impedances, the T\&R coefficients of each STAR element can be controlled in terms of the amplitude and phase-shift responses. 
In \cite{9570143}, three practical operating protocols for STAR-RISs are proposed, namely, energy splitting (ES), mode switching (MS), and time switching (TS).
Moreover, for STARS with passive-lossless elements, the phase shifts of the T\&R coefficients are coupled. In \cite{xu_correlated}, a correlated T\&R phase-shift model was proposed for STARS.

\begin{figure}[t!]
    \centering
    \captionsetup{font={footnotesize}}
    \begin{subfigure}[t]{0.24\textwidth}
        \centering
        \includegraphics[width=1\textwidth]{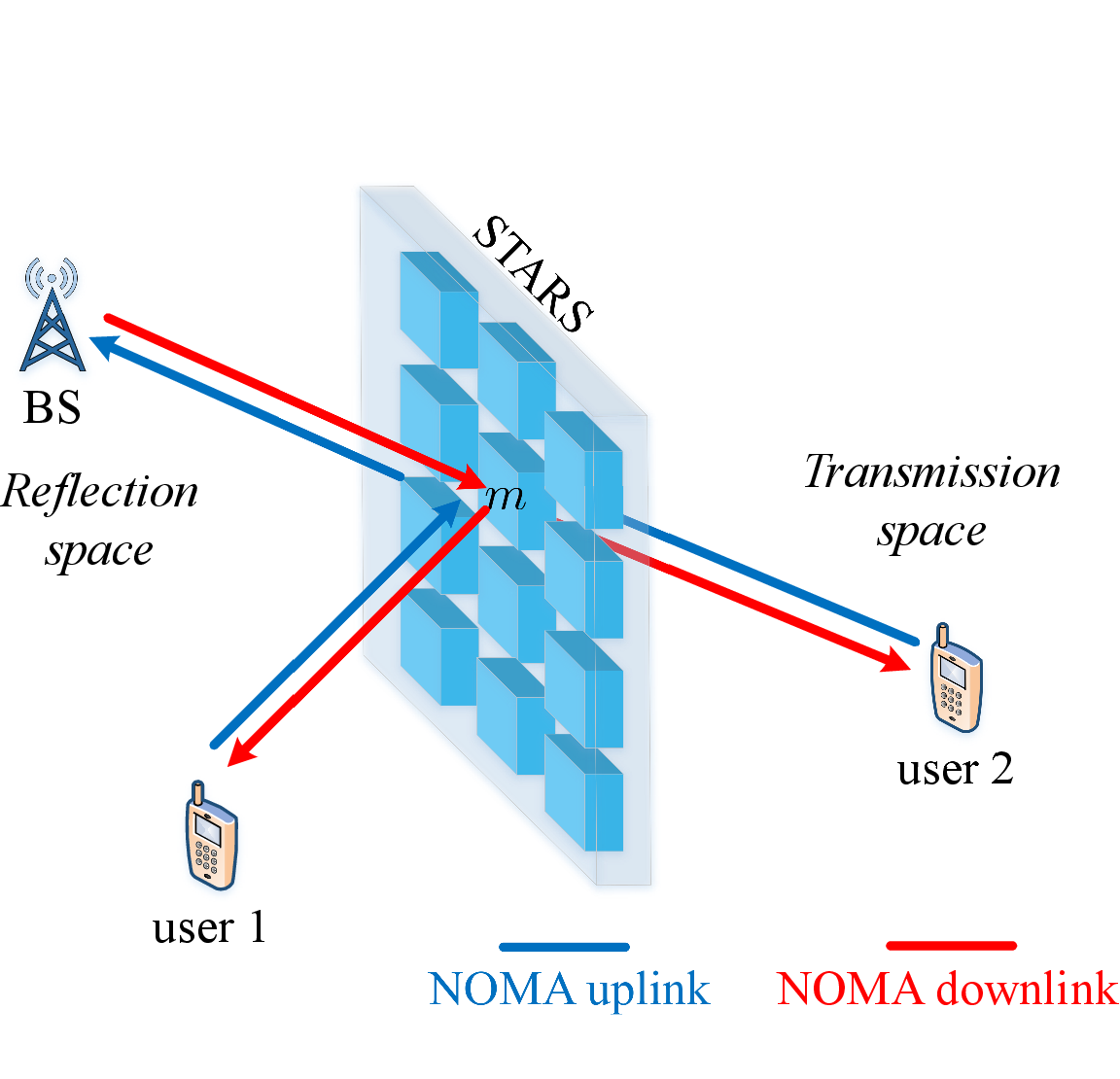}
        \caption{Network illustration.}
        \label{fig:STARS-NOMA}
    \end{subfigure}
    \begin{subfigure}[t]{0.24\textwidth}
        \centering
        \includegraphics[width=1\textwidth]{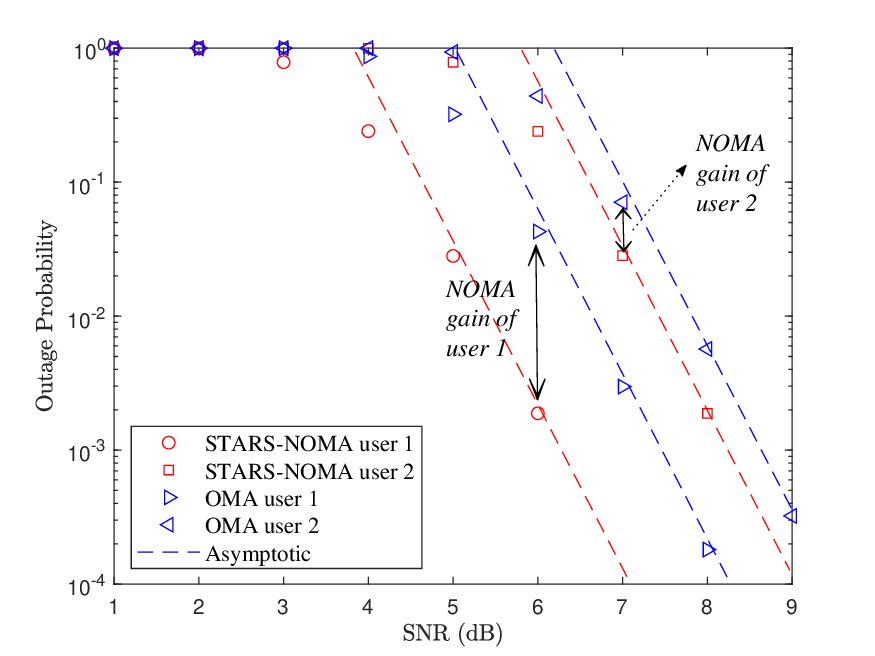}
        \caption{Outage probabilities.}
        \label{fig:outage}
    \end{subfigure}
    \caption{STARS-NOMA network and outage probabilities for NOMA users.}
\end{figure}

\subsubsection{STARS enhanced NOMA}
Compared to conventional transmitting/reflecting-only RISs, STARS provides more DoFs for facilitating the T\&R NOMA. Recalling the fact that the performance gain of NOMA over OMA relies on the channel condition disparity between paired users. The key idea of T\&R NOMA is to pair a transmitted user and a reflected user together for NOMA transmission, as shown in Fig.~\ref{fig:STARS-NOMA}. The advantage of T\&R NOMA is that by adjusting the T\&R amplitude and phase-shift coefficients, a distinct channel difference can be achieved between users located at each side of STARS, which leads to high NOMA performance gain. Such a T\&R NOMA operation addresses the issue when employing transmitting/reflecting-only RISs in NOMA, where the users surrounding the transmitting/reflecting-only RIS might have a similar channel condition. To further illustrate the benefits of T\&R NOMA, Fig.~\ref{fig:outage} depicts the outage probabilities for STAR-NOMA and STAR-OMA users \cite{xu_correlated}. It can be observed that with the aid of STARS, NOMA outperforms OMA for both users' outage probabilities. Moreover, the performance gain of STAR-NOMA over STAR-OMA is more pronounced in the high signal-to-noise ratio (SNR) region.

\subsubsection{Existing Works on STARS-NOMA} Motivated by the potential benefits brought by STARS for NOMA, increasing research interests have been devoted to the performance analysis and optimization of STARS-NOMA.  
The STARS-NOMA system was first proposed in \cite{9685891} where the achievable sum rate was maximized by jointly optimizing the decoding order, power allocation coefficients, active beamforming, T\&R beamforming. 
In terms of performance analysis of the STAR-NOMA network, various studies have been conducted to evaluate the bit error rate (BER) performance~\cite{9786807}, the impact of the correlated T\&R phase-shift~\cite{xu_correlated}, the ergodic rate for users over Rician fading channels~\cite{9856598}, the outage probability of users over spatially correlated channels~\cite{9774334}, and the performance of an uplink STARS-NOMA network~\cite{9935303}. 
Apart from these studies, recent works focused on STARS-NOMA networks with the more complex distribution of users and scatters in the environment. For example, In \cite{9808307}, a fitting method was proposed to model the distribution of STARS-NOMA channels composing small-scale fading power as the tractable Gamma distribution. Based on the proposed model, a unified analytical framework based on stochastic geometry was provided to capture the random locations of STARS, BSs, and users. In \cite{9843866}, the ergodic rate of the STARS-NOMA system was investigated, where the direct links from the BS to cell-edge users are non-line-of-sight due to obstacles, and STARS is used to provide line-of-sight links to these cell-edge users.

In terms of optimization, existing works focused on the beamforming design and power allocation scheme for STARS-NOMA. In \cite{9863732}, the framework for applying NOMA to STARS networks was first proposed and a cluster-based beamforming design was adopted to jointly optimize the decoding order, power allocation coefficients, and beamforming at both the BS side and STAR-RIS side. In \cite{9847399}, a joint power and discrete amplitude allocation scheme was proposed for the STARS-NOMA system, which can reduce CE workload and hardware complexity.
In addition, recall that one of the benefits of STARS is that it can be seamlessly integrated with other communication technologies. Exploiting STARS with other technologies is a major interest for the research community. For example, in \cite{9956998}, the concept of index modulation (IM) was first incorporated into the STARS-NOMA system to improve spectral efficiency. Specifically, the proposed IM-aided STARS-NOMA system enables extra information bits to be transmitted by allocating subsurfaces to different users in a pre-defined subsurface allocation pattern. In \cite{10163896,9961851}, a new active STARS-NOMA network was investigated. In particular, simple active devices are introduced into the STARS to overcome the “double-fading” effect. Furthermore, STARS was also exploited in a MIMO-enabled NOMA system~\cite{9956827}. Among others, exploiting STARS-NOMA in secured communication is an appealing option. In \cite{xie2023physical,xu2022starPLS}, the secrecy performance of STARS-NOMA systems over general cascaded $\kappa-\mu$ fading and Rayleigh fading channels was investigated.

\begin{figure}[t!]
    \begin{center}
        \includegraphics[scale=0.38]{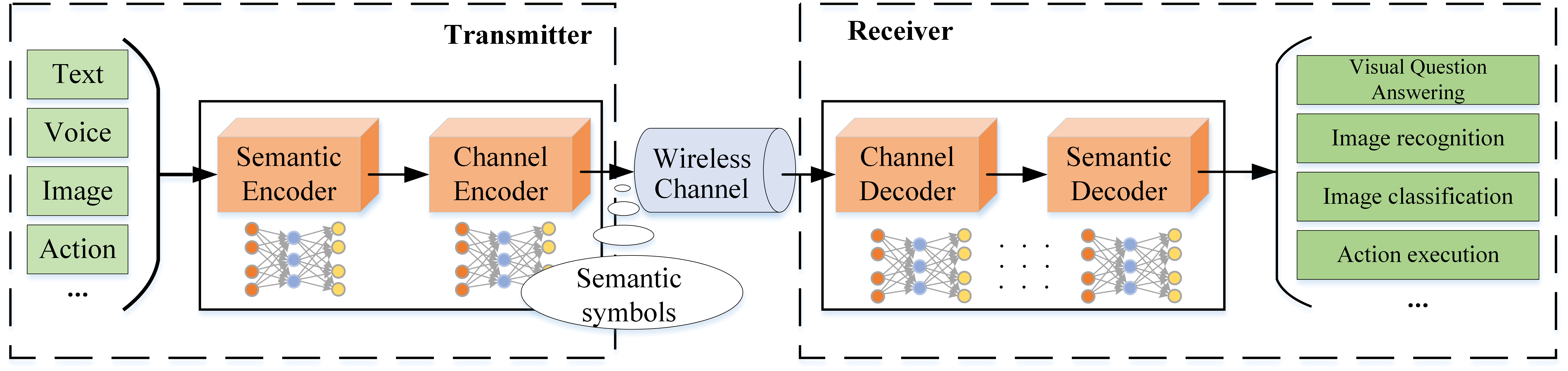}
        \caption{Illustration of an end-to-end semantic communication framework.}
        \label{SemCom}
    \end{center}
\end{figure}

\subsection{Interplay between NOMA and Semantic Communications}
The Shannon classical information theory guides the development of communication systems from 1G to 5G. Thanks to the great efforts of information and communication technology (ICT) researchers in the past few decades, many efficient communication technologies were proposed to address the technical-level communication problem~\cite{Shannon2} and thus approach the fundamental Shannon limit. However, the communication requirements in the upcoming 6G, such as extremely high data rate and massive connectivity, are far from being satisfied, which calls for a new information transmission paradigm to be developed. In this subsection, we first provide a brief overview of semantic communications and then discuss the interplay between semantic communications and NOMA.

\subsubsection{Overview of Semantic Communications} Compared to conventional information transmission, which treats the source data equally into the bit sequences and ignores the intrinsic meaning, semantic communications are task- and goal-oriented. As shown in Fig. \ref{SemCom}, in semantic communications, only the information (conveyed by semantic symbols) which is related to the specific meaning/actions/goals required by the receiver should be transmitted~\cite{qin2021semantic,10328187}. This is achieved by the semantic and channel encoders at the transmitter, which is empowered by advanced machine learning tools, e.g., deep learning, for the semantic feature extraction from the source data (e.g., text, voice, image, etc.). When receiving the semantic symbols through the wireless channel, the receiver employs the channel and semantic decoders to accomplish the required tasks and/or achieve the transmission goals (e.g., visual question answering~\cite{guo2024videoqascadaptivesemanticcommunication}, image recognition/classification \cite{8723589,9398576,guo2024digitalscdigitalsemanticcommunication}, action execution \cite{9796572}, CSI acquisition~\cite{9954153,guo2024deeplearningcsifeedback,guo2024deepjointcsifeedback}, molecular communications \cite{10606344}, etc.). It can be observed that only the key information needed to perform the function or deliver the meaning needs to be transmitted in semantic communications, the required radio resources can be greatly reduced. Therefore, semantic communications are more energy-efficient and sustainable than conventional bit-level communications. More importantly, existing research contributions showed that the superiority of semantic communications over conventional bit-level communications is more significant in bad wireless conditions, which enhances the reliability of communication systems. It can be observed that semantic communications provides a promising information transmission option for future wireless networks. In the following, we focus our attention on the interplay between NOMA and semantic communications in multi-user networks, namely, NOMA-enabled semantic communications and semantic communications-enhanced NOMA.

\subsubsection{NOMA-enabled Semantic Communications} On the one hand, when employing semantic communications in multi-user networks, one of the most fundamental problems is how to develop efficient multiple access schemes to accommodate the new semantic users given the limited radio resources. As a first step, the authors of \cite{xidong} studied the NOMA design for facilitating the heterogeneous semantic and bit multi-user communication, where a transmitter simultaneously serves one semantic user and one conventional bit user. Recalling the fact that the transceivers of semantic communications have to be jointly trained in advance, it is almost impossible for the conventional bit user to employ SIC to detect the semantic user's signal, which leads to a fixed \emph{bits-to-semantics SIC ordering} in the heterogeneous transmission. To manage the interference received by the bit user and maximize the resource efficiency, a novel semi-NOMA scheme was proposed in \cite{xidong}. The total available bandwidth is divided into one non-orthogonal sub-band for the heterogeneous transmission and one orthogonal sub-band for the bit-only transmission. By adjusting the bandwidth and power allocation at the transmitter, the proposed semi-NOMA can unify both conventional NOMA and OMA schemes and thus achieves the maximum semantic-versus bit rate region, i.e., being the optimal multiple access scheme for the considered heterogeneous semantic and bit multi-user communication. Moreover, the authors of \cite{10225385} explored the employment of NOMA to support pure semantic multi-user communication, which is capable of serving multiple users with different modalities of data via semantic communications. The results showed that the proposed NOMA-enabled semantic multi-user communication has strong robustness and can achieve higher spectral and power efficiency.

\subsubsection{Semantic Communications-enhanced NOMA} 
\begin{figure}[t!]
    \begin{center}
        \includegraphics[scale=0.6]{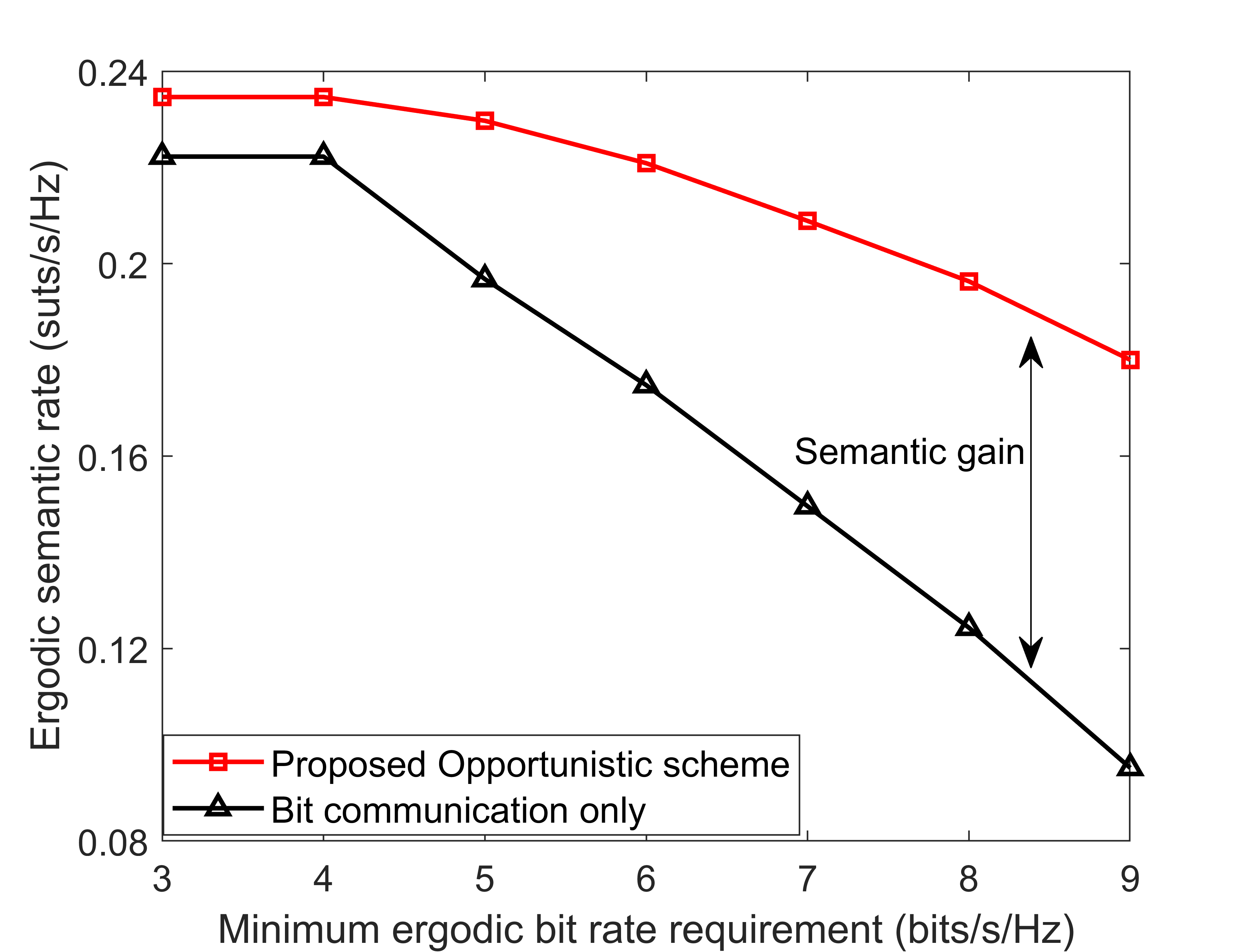}
        \caption{The ergodic (equivalent) semantic rate achieved by the secondary NOMA user versus the ergodic bit rate required by the primary NOMA user over fading channels. The system parameter settings can be found in \cite{Xidong2}.}
        \label{SemCom_Result}
    \end{center}
\end{figure}

On the other hand, how to employ semantic communications to enhance the NOMA performance is another interesting research topic. One important observation from existing research is that for achieving the same transmission goal, semantic communications generally require fewer radio resources (e.g., transmit power and bandwidth) than conventional bit-level communications. Motivated by this, an opportunistic semantic and bit communication strategy was proposed in \cite{Xidong2} for the secondary user in uplink NOMA. The key idea is that the secondary user can appropriately select a semantic or bit communication option to control the resulting co-channel interference when reusing the resource block of the primary NOMA user. The advantages of the proposed opportunistic semantic and bit communication strategy can be explained as follows. When the primary NOMA user has a high communication requirement (e.g., a high-resolution video user), the transmit power that can be used by the secondary NOMA user is strictly capped. In this case, employing bit-level communications at the secondary NOMA user cannot achieve a satisfactory communication performance due to the limited power budget. As a result, semantic communications comes to the rescue, which guarantees the performance of the secondary user in the low SNR regime. Otherwise, when the primary NOMA user has a low communication requirement (e.g., an IoT user), the secondary NOMA user can use the bit-level communication option to achieve the best performance using a sufficiently high power (i.e., high SNR regime). Fig. \ref{SemCom_Result} provides simulation results to further demonstrate the benefits of the proposed opportunistic semantic and bit communication strategy. It can be observed that with the employment of semantic communications, the performance of the secondary NOMA user can be greatly improved, especially when the communication requirement of the primary NOMA user is high.

\section{Towards Learning-Based Next Generation Multiple Access: Challenges and Future Work}

Model-driven methods take advantage of various domain knowledge in the problems of NGMA. However, the model-driven methods usually suffer from high computational complexity and are restricted by the modeling of domain knowledge. As a feasible alternative, data-driven methods learn from data to solve detection and estimation problems, providing a new way to improve performance and reduce computational complexity. The use of learning-based methods for NGMA still requires a lot of in-depth exploration. Some of the challenges and future work are discussed in this section.

\subsection{Construction of Training Dataset}
The foundation of data-driven methods is the source of training data. The performance of data-driven methods is highly dependent on the consistency between the training and testing sets. Most existing data-driven methods for NGMA use data generated according to some channel models in both training and evaluation. Given the historical CSI data accumulated in BS, it is possible to customize the algorithms for a specific cell by using real dataset of the cell. In situations where real data cannot be obtained, ray tracing could be employed to generate more realistic channel data. Other features beyond channel characteristics would be exploited in learning-based methods. For example, data-driven methods can learn the access probability of users, and use the prior information to enhance random access. 

If only a small amount of real data is available, a potential approach is to use simulated data to train the learning model, and then use transfer learning to fine-tune the learning model on the real data. In practice, it is beneficial to use newly obtained data to update the model. Another way is to generate the dataset using generative AI methods, such as VAE and generative adversarial network. 

\subsection{Specialized Neural Network Design}
The convolutional neural network explores structural information in images, and the long short-term memory network explores the relevance of sentences in natural language. Therefore, it is necessary to design specialized networks for different tasks in communication systems. In NGMA, domain knowledge should be exploited in the design of learning-based methods. A potential solution is to use deep unfolding methods, which construct the neural network according to some iterative optimization algorithm. The resulting neural network inherits the domain knowledge embedded in the traditional optimization problem.

Furthermore, computational complexity and storage requirement of the learning-based methods are strictly restricted in our tasks. Light-weight networks are desired, especially for low-cost devices. One approach to speed up neural network deployments, while reducing model storage size, is model pruning. Other techniques for model compression, e.g., quantization, knowledge distillation, and low-rank factorization, would also be useful tools in building the specialized neural networks for our task. 

\subsection{Scalability and Generalization}
Scalability and generalization are major concerns in the implementation of learning-based methods for NGMA. The existing work on machine learning for NGMA has shown reduced computational complexity in the inference stage, in comparison to the traditional optimization-based iterative methods. However, the training phase of learning-based methods is time-consuming. Given various communication scenarios, e.g., different environments, SNRs, number of active users and so on, applying one learning model for one specific scenario is costly in both model training and model storage. To relieve the burden in training a number of models, one could employ meta-learning, which is a process that helps models learn new and unseen tasks on their own with little effort. Furthermore, deviations between training and testing data can lead to a decrease in performance. For example, a model trained at one SNR may experience performance degradation at another SNR. Mixing training data with different settings may improve the robustness of data-driven methods. However, the ability of the network is limited, which can also lead to unsatisfactory training results. It would be interesting to investigate the mechanisms that can adjust the learning model to adapt to different scenarios.

\section{Conclusion}
In this article, we have presented a comprehensive overview of the research efforts to date in signal processing and learning for NGMA, with a focus on MRA and NOMA. In particular, the fundamental limits and practical schemes are reviewed first, and then the state-of-the-art research contributions in the use of advanced signal processing and machine learning techniques are provided. We further discuss the promising interplay between NOMA and other new next-generation technologies including NFC, ISAC, STARS, and semantic communications. Finally, some challenges and future work in the use of learning-based methods for NGMA are listed. We believe that the in-depth overview of intelligent signal processing and learning for NGMA will provide inspiration for the development of new solutions for NGMA.


%
%

\ifCLASSOPTIONcaptionsoff
  \newpage
\fi



%

\bibliographystyle{IEEEtran}
\bibliography{IEEEabrv,bib_paper}

\end{document}